\documentclass[aps,prx,twocolumn,showkey,square,numbers,amssymb,amsmath,longbibliography,nobibnotes,nofootinbib]{revtex4-2}

\usepackage[utf8]{inputenc}
\usepackage[T1]{fontenc}
\usepackage[colorlinks=true,linkcolor=red]{hyperref}
\usepackage{verbatim}
\usepackage{todonotes}
\usepackage{soul}

\usepackage{graphicx}
\usepackage{float}
\usepackage{epsfig}
\usepackage[font=small]{caption}
\usepackage{subcaption}
\usepackage[shortlabels]{enumitem}
\usepackage{cancel}
\usepackage[left=1.6cm,right=1.6cm,top=2cm,includefoot,bottom=1.5cm]{geometry}

\usepackage{bm}
\usepackage{dsfont}

\usepackage{amsmath}
\usepackage{mathtools}
\usepackage{hyperref}
\usepackage{epsfig}
\usepackage{eucal}
\usepackage{graphicx}
\usepackage{float}
\usepackage[utf8]{inputenc}
\usepackage{caption}
\usepackage{color}

\newcommand{\bsigma}{{\bm{\sigma}}}
\newcommand{\bI}{{\bf I}}

\begin{document}

\title{Price of information in games of chance: a statistical physics approach}
\author{Luca Gamberi, Alessia Annibale, Pierpaolo Vivo}

\affiliation{Quantitative and Digital Law Lab, Department of Mathematics, King's College London,
	Strand, WC2R 2LS, London (United Kingdom)}


\newcommand{\R}{\mathbb{R}}  
\newcommand{\C}{\mathbb{C}}  
\newcommand{\diff}{\textup{d}} 
\newcommand{\Tr}{\textup{Tr}} 
\newcommand{\conj}[1]{\overline{#1}} 
\newcommand{\diag}{\textup{diag}} 
\newcommand{\Mean}{\mathbb{E}} 
\newcommand{\prob}[1]{\mathbb{P}\left\{#1\right\}} 
\newcommand{\EvalAt}[1]{\Bigg\vert_{#1}} 
\newcommand{\Eye}{\mathds{1}} 
\newcommand\indices[1]{\langle #1 \rangle} 


\setlength{\skip\footins}{1.5cm}

\begin{abstract}

Information in the form of \emph{data}, which can be stored and transferred between users, can be viewed as an intangible commodity, which can be traded in exchange for money. Determining the fair price at which a string of data should be traded is an important and open problem in many settings. In this work we develop a statistical physics framework that allows to determine analytically the fair price of information exchanged between players in a game of chance. For definiteness, we
  consider a game where $N$ players bet on the binary outcome of a stochastic process and share the entry fees pot if successful. We assume that one player holds information about past outcomes of the game, which they may either use exclusively to improve their betting strategy or offer to sell to another player. We find a sharp transition as the number of players $N$ is tuned across a critical value, between a phase where the transaction is always profitable for the seller and one where it may not be. In both phases, different regimes are possible, depending on the ``quality'' of information being put up for sale: we observe \emph{symbiotic} regimes, where both parties collude effectively to rig the game in their favor, \emph{competitive} regimes, where the transaction is unappealing to the data holder as it overly favors a competitor for scarce resources, and even \emph{prey-predator} regimes, where an exploitative data holder could be giving away bad-quality data to undercut a competitor. Our analytical framework can be generalized to more complex settings and constitutes a flexible tool to address the rich and timely problem of pricing information in games of chance.
\end{abstract}

\maketitle
\section{Introduction}

Since Maxwell conceived his demon thought experiment \cite{Rex17,Maruy09}, information has played an essential role in shaping our understanding of the statistics and thermodynamics of physical systems. In the enigma formulated by Maxwell, a demon living in an isolated system would be able to let fast-moving molecules through a mass-less door leaving the slow-moving ones isolated in one compartment, in apparent violation of the second law of thermodynamics. However, the demon would be required to ``know'' the velocities or energies of individual gas molecules, and selectively allow them to pass through the barrier between two compartments. More recently, Landauer's principle \cite{landauer61,bennet03,sagawa09} shed further light on the energy costs of information manipulation,
establishing a lower bound (temperature-dependent) on the energy needed to erase one bit of information stored in a computer. Meanwhile, Shannon \cite{shannon48} had already startled the scientific community by proposing his mathematical theory of information, which forms the theoretical basis of data compression. In more recent years, the field of ``information thermodynamics'' has flourished \cite{sagawa09,parrondo15}, with a number of spectacular results where information content and thermodynamic variables are treated on an equal footing.

Information has had a profound impact on modern economic theory and policy as well as physics, especially in the context of competitive equilibrium analysis and efficient market hypothesis \cite{stiglitz02,berg02,seoane21}. With the dawn of the Big Data era, another incarnation of ``information'' -- in the form of large amounts of data, which can be stored and transferred between users -- started to become apparent: that of a \emph{commodity}, which can be traded in exchange for money. Examples include genome sequences, financial time series, software, and digital music records, among many others: the implications of our ability to attach a price tag to intangible entities -- made up of bits, stored and transmitted electronically -- are far-reaching, and have already had a profound impact on the way modern economies operate.

Contrary to the traditional ``supply/demand'' equilibrium, though, which forms the theoretical framework to price \emph{material} commodities \cite{linde09,gu11}, intangible assets feature some very peculiar traits, which make the question ``how much should I reasonably pay for this digital item?'' so much harder in this context.

First, the costs for duplication and distribution of data are typically much lower than for material assets. Moreover, at odds with material goods whose scarcity normally drives the price up, intangible assets like software, news, and music may actually become more valuable the more widespread they are: this often leads to extremely low-priced promotional offers, as a wide circulation and ``network effects'' of a product are considered more valuable than the immediate loss of revenues they would entail. Also, information goods lend themselves to aggressive price discrimination strategies, whereby different levels of pricing can be easily implemented to meet the demands of different customer profiles: typical strategies include \emph{windowing} -- bringing information goods like a movie or a book in different forms, at varying times on to the market -- \emph{versioning} -- the offer of a digital good in different versions and with different privileges, allowing the customers to select their preferred level of engagement -- and \emph{bundling} -- whereby two or more goods are combined and offered together on the market for a more convenient price \cite{ahituv80,linde09,Pei19}.  Furthermore, information goods sale and usage is \emph{non-rival}, i.e. their consumption by one user does not preclude other sales or compromise their quality. This should be contrasted with physical assets for which property often implies exclusivity and their usage implies depreciation \cite{tonetti20}.

Information assets can be categorized into digital and data assets, which share similar pricing challenges but differ in their consumption. Digital assets like e-books, movies, and software are designed for specific purposes, whereas data asset consumption use cases may be known only to the buyer. Data assets exhibit less frequent versioning because they can be aggregated in various ways based on the specific use case and data analytics needs. Additionally, data retains value even when segmented, while disaggregating digital goods often renders them worthless. In essence, data can be viewed as a ``raw'' commodity in contrast to the more refined nature of digital goods \cite{Pei19}.

The problem of data pricing comes in a huge variety of potential settings, with several contributing factors and competing interests to take into account in any particular situation. In this work, 
we focus on a specific
class of problems, known as ``games of chance'', i.e. games where different agents -- holding on to different levels of information -- compete for a pot of money by betting on the outcome of a stochastic process.

In this context, we propose 
a general and flexible framework grounded in statistical physics to determine analytically the \emph{fair price of information}. For simplicity, we further analyze in detail a simple game where $N$ players pay a fee $\phi$ to place a bet on the binary outcome $\{+1,-1\}$ of an underlying stochastic process and share the entry fees pot if their bet is successful. One of the players (the \emph{data holder}) is special, though: they hold a long sequence of past outcomes of the game, which they can decide to either use exclusively -- to improve their  forecast on future outcomes of the game and therefore of their own winning chances -- or to partially share with another player (the prospective \emph{buyer}) in exchange for a monetary compensation.

There are several real-life scenarios in which agents compete using their resources -- capital and information -- and achieve gains often at the expense of others' losses. For instance, in trading, hedge funds that have developed proprietary algorithms may decide to sell their predictive models (or the information they generate) to other traders for immediate revenue gains, but at the risk of increasing correlations between investment strategies that may potentially lead to higher stock prices. Similarly, on a smaller market scale, retailers with detailed customer behavior insights may sell their datasets to competitors, or private equity firms can share market information while still competing to acquire shares of the same companies. All these real-life scenarios mirror -- at least conceptually -- the ``game of chance'' setting we present here, where an informed player may decide whether to sell their knowledge or hold on to it, balancing immediate profit against the dilution of their competitive advantage.

In our simplified setting, we are able to compute analytically the fair price range the data holder should put their information up for sale. We find that the ``ecosystem'' of players displays a rich and non-trivial \emph{informational landscape} depending on the quality of information being potentially traded as well as the number of participants in the game. First, we observe a sharp transition in the number of players $N$ participating in the game, which separates a phase where a crowded game makes the transaction always appealing and favorable to the data holder from another where the scarcity of resources (namely, the fees contributed by the $N-2$ uninformed players) may lead to the onset of a fiercely \emph{competitive} regime, where information sharing is discouraged. A similar phase transition in the number of information-processing agents was observed in \cite{berg02} when considering financial market efficiency.

Depending on the parameters of the data being offered, other situations are possible, e.g. a \emph{symbiotic} regime where the transaction benefits both the seller and the buyer at the expense of the $N-2$ uninformed players, as well as a \emph{``prey-predator''} regime, where the quality of data being offered is low enough that it would actually mislead the buyer into betting on the ``wrong'' outcome, with clear benefits for a potentially exploitative seller in terms of thwarted competition. Our setting resembles a small ecological network \cite{allesina12} in which the interplay between scarcity of resources, competition, and the entry cost threshold creates interesting consequences on the information trading process.

The advantage of using a statistical physics approach to the game is threefold: first, we can apply a standard ``expected utility'' maximization argument \cite{bergemann18} to compute analytically the \emph{ex-ante} ask/bid price curves of the seller and the buyer such that the game 
becomes more profitable for one or both. Second, the framework is sufficiently flexible and general that it can be adapted to more complex and realistic cases. Third, the problem -- even in this very simple setting -- turns out to be extremely rich and non-trivial, with phase transitions and regime changes also observed, for instance, in the context of information aggregation for achieving consensus \cite{livan13} or in the processing of financial market information \cite{berg02}. 

The main pricing mechanism, when the transaction goes through, lies in the interplay between \emph{increased competition} -- namely, the fact that the seller expects to realize a lower gain from the game, due to a larger expected number of winners -- and a \emph{reduced risk} -- because the seller will realize a fixed, non-random income (the payment made by the buyer to obtain the data). Quite interestingly, the financial goals of seller and buyer may be aligned or misaligned, depending on the regions in parameter space and the total number of other players, giving rise to a complex landscape of interactions rooted in the \emph{informational asymmetry} \cite{ichihashi22,gradwohl23} of the game. 

Our work does not delve into the discussion of so-called \emph{disclosure rules}, i.e. the mechanisms by which sellers and buyers share some information about the assets to reach an agreement on the price, which is typically the focus of economic studies. Typically, sellers can reveal information prior to the sale \cite{babaioff12} or they can commit to a sale mechanism before any state variable becomes known \cite{bergemann18}. In our study, we do not establish any specific disclosure policy and we mainly assume that the data holder is a reputable player, who is not interested in over-exploiting the prospective buyer but simply wants to act according to fair market dynamics.

The plan of the paper is as follows. In Section \ref{sec:related} we briefly summarize the existing literature (mainly in economics) dealing with valuation and pricing of intangible commodities, and point out the differences with our setting. In Section \ref{sec:betting}, we develop a general statistical physics framework, which we specialize to a binary-outcome game in subsection \ref{IIIa}. We then cast the problem of determining the impact on players' wealth when information is exchanged into establishing the ``fair'' price range for the trade of such information in subsection \ref{sec:fair}, before a brief recap of the main ingredients of the general framework in subsection \ref{sec:int_summary}. In Section \ref{sec:summary}, we summarize our findings for a few specific regions in parameter space, and we discuss the ``pricing phase transitions'' we observe. In Section \ref{sec:baseline}, we compute the baselines (expected wealth change) for the prospective buyer -- when they place a bet in the absence of any extra information about the game or the other players' strategy -- and the prospective seller -- who will anyway still use the data they hold exclusively to gain an edge over the other uninformed players. In Section \ref{sec:expectedafter}, we compute the expected wealth changes of the buyer and the seller after the transaction has been arranged, which are necessary ingredients to compute the price curves. In Section \ref{sec:conclusion}, we offer some concluding remarks and an outlook on future research and interesting extensions that can be tackled using our framework. The Appendices are devoted to more technical remarks.

\section{Related Works}\label{sec:related}

The intriguing features of digital and data goods and their pricing problem have attracted the attention of several communities. Theoretical economists have significantly contributed to the understanding of probabilistic theories of information exchange. This is particularly true in the study of cost functions and the development of digital product menus that are optimally priced in monopolistic contexts, as explored in \cite{admati86}.

The precursory methodology introduced by \cite{wald45} on dynamic decision-making processes, specifically the sequential probability ratio test, establishes a rule for an observer to decide whether to prolong or terminate the observation of a valuable state variable. Building upon this, \cite{girshick49} incorporated the consideration of costs linked to various strategies and sampling efforts, particularly on the trade-off faced by observers when balancing the value against the cost implications of their observations.

Further research has focused on the costs associated with information acquisition. The study \cite{zhong20} introduces a comprehensive theory on the cost for decision-makers to sequentially acquire information, providing insights into what are deemed ``reasonable'' cost functions, especially for Bayesian decision-makers faced with single-instance decisions and gathering data through the realisation of Blackwell experiments, i.e. abstract mathematical information structures that produce an observation of a state of nature according to a predefined probability, unknown to the observer. Alternatively, papers such as \cite{leahy22} have focused on Shannon information-based cost models, designing entropy-based cost functions and examining how these functions correlate with observable ``willingness-to-pay'' data.

Other theoretical investigations have grappled with more realistic scenarios. For instance, \cite{sherman49} incorporates game theory concepts like reconnaissance to evaluate a game's value based on information gained -- from other players -- through strategic moves. The work in \cite{deep17} addresses the pricing of information queries from databases and analyzes the arbitrage-free condition in the application. Game theorists have also looked at the topic in some depth, with \cite{myerson84,khouzani18} looking at cooperative games.

In the field of rational inattention and information costs, in which an agent purposefully makes decisions based on incomplete information as acquiring complete knowledge would be prohibitively costly, the studies carried out in \cite{denti22,rustichini22} have contributed most significantly, among others. The former provides a testable framework for posterior separability in cost functions, while the latter examines the relationship between information cost and the prior beliefs of the decision-maker, highlighting certain incompatible assumptions with the defined cost function. The impact of marginal costs is explored in \cite{tamuz23}.
A recent study by \cite{nouripour24} introduces a competition dynamics among symmetric buyers in an information market, investigating the characteristics of the information cost function for binary choices and the strategies for equilibrium information allocation, considering factors like seller recommendations, buyer obedience and the accuracy of other buyers' revealed preference. In \cite{ichihashi21}, the focus shifts to the impact of data externalities among data owners with overlapping information and their influence on information costs. Other approaches seen in \cite{li13,bergemann15,babaioff12,andreas22, gradwohl23} study the compensation mechanism for selling third-party privacy-sensitive information queries, data sharing, and the losses due to voluntary data disclosure.

The challenge of a practical approach to data pricing and valuation has become a cornerstone in the digital economy, giving birth to a new cross-disciplinary research field dubbed \emph{infonomics}, still in its infancy \cite{laney17}, primarily stemming from business economics. This emerging area aims to explore the topic in a more pragmatic and empirical way, particularly focusing on the characteristics of data and digital assets (e.g., pricing of non-rival goods, and pricing with bundling or versioning), their roles in the economy, and their contributions to value creation for both companies and public institutions \cite{viswanathan05,Azcoitia21,Fleckenstein23,Coyle22,kannan18,Pei19}. In parallel with the rising business research, many companies and startups have been founded in the past few years to provide data valuation services: in the absence of an established practice, though -- or even common goals -- they seem to rely mostly on general heuristics and bespoke or proprietary models, which are claimed to work only in the specific settings they were developed \emph{ad hoc} for.

Computer scientists have explored algorithmic approaches to data valuation in which the pay-off is usually measured by the impact of information on a prescriptive action. In \cite{agarwal19}, a number of algorithmic strategies are proposed for pricing training data designed to improve a machine learning task. In \cite{GhorbaniZou19}, the authors introduced a ``data Shapley'' metric for valuing individual data points in machine learning, showing why such value metric can be used effectively in valuation tasks that depend on the quality of information being used. 

In the physics literature, examples of valuation are very limited. Some works have explored the consequences of rational agents in a non-ergodic market. In \cite{roger20}, the authors conceptualize a ``quasi-ergodic'' market where agents continuously update their beliefs in response to observed outcomes. Despite agents having divergent beliefs and engaging in trades, the model suggests that they cannot derive learning or achieve consensus from the market dynamics. The model does not incorporate the costs of acquiring information, focusing solely on a portfolio optimization problem involving just two securities. Another study \cite{scarani23}, which focuses on applied expected utility theory to asset trading, draws parallels between rational players' information strategies and stochastic thermodynamics.  Unique to this model is the introduction of an ``energy expenditure'' required for participation in the trading game. The study examines the behaviors of these agents who have the option to enter the game employing either risk-prone or risk-averse fee payment strategies. 
In the ``poker game'' examined in \cite{javarone15}, players are categorized as either rational or irrational, where the rational players utilize the information from the cards revealed to them, while the irrational players do not. The study reveals that the effectiveness of a strategy, and consequently whether the processed information is ``valuable'', is contingent on the presence of a sufficient number of rational players. If the number of rational players is too low, the game becomes a pure game of chance, regardless of the information available. Interestingly, the authors observe a transition between two states - strategy-driven and random - as the ratio of rational to irrational players varies. Finally, in \cite{dinis04}, the authors show that random sequences of two losing games may result into a winning game. This phenomenon is modeled on the theory of Brownian ratchets, systems known for their ability to mitigate thermal fluctuations. A key finding of this research is the counter-intuitive role of information in this context. The study demonstrates that in such systems, using information about the previous outcome to adjust one's gambling strategy may actually be less effective than it appears and may be outperformed by purely random choices.

However, these studies do not consider the impact of information sharing on the pay-off for the parties involved, nor do they explore the dynamics of their interactions. We maintain that this presents a significant challenge, well-suited for examination by the statistical physics and complex systems communities. Our research is designed to provide the foundational setting for investigating the phenomena of information sharing through the lens of statistical physics, offering an alternative approach to traditional theoretical economics frameworks. This perspective not only broadens the scope of 
those communities' interests,
but also will enable new ways of tackling information pricing issues.

\section{Statistical Physics of Information-Trading Games}\label{sec:betting}

To set up the statistical physics framework, we will consider a stochastic process that returns a value $\sigma_{t}\in \mathcal{A}$ at time $t$, and we will denote with 
$\sigma_{0,\ldots,t-1}$ a particular realization of its trajectory from time $0$ to $t-1$.
This could be the sequence of $t$ independent coin tosses ($\mathcal{A}={\pm 1}$), or a Markov chain of length $t$ over $K$ states ($\mathcal{A}={1,\ldots,K}$), or any other (arbitrarily correlated and complex) process.

Suppose that $N$ agents, referred to as \textit{players}, can enter up to $M$ rounds of a game of chance, built on the outcomes of this underlying stochastic process, by paying a fee $\phi$ in exchange to the ability to place one bet per round and realize a profit, should they be able to predict the outcome of the game. 
The players are ordinarily unaware of the full underlying stochastic process that generates the outcomes $\sigma_{t}$, but may hold some partial information about it (for example, a portion of trajectory of the $R$ past outcomes, $\sigma_{t-R,\ldots,t-1}$). The state space $\mathcal{A}$ is revealed in advance to all players that wish to participate.

In particular, before the $t$-th outcome is announced, each player $j$ ($j=1,\ldots,N$) will have committed $\phi$ money units to a common pot, and will have placed their bet $\sigma_t^{(j)}\in \mathcal{A}$ on the outcome of the game. The game's \emph{configuration space} at round $t$ is given by the vector $\bm\sigma_{t}=(\sigma^{(1)}_{t},\ldots,\sigma^{(N)}_{t})$, comprising the list of all bets placed at that round.
We assume that the players start placing their bets at $t=0$ on the values of the future states $\sigma_0,\ldots, \sigma_{M-1}$ of the process, using information they may hold on past outcomes. Thus, the game is fully specified by the joint distribution of the $M$ sets of bets and $M$ outcomes (conditional on the past) 
\begin{align}
& P(\sigma_{0,\ldots,M-1}, \bsigma_0, \ldots, \bsigma_{M-1}|\sigma_{-R,\ldots,-1})
\nonumber\\
&=P(\sigma_{0,\ldots, M-1}|\sigma_{-R,\ldots,-1}) 
\prod_{t=0}^{M-1} P(\bsigma_t|\sigma_{-R,\ldots,t-1})\ .
\label{eq:joint}
\end{align}
In any realistic setting, such distribution is 
unknown to the players, who will try 
to estimate it based on the information they have. 

In the most general scenario, not all players  
will have the same information about past outcomes: at each round, each player $i$ will hold a different piece of information and will estimate (or will make assumptions on) the information held by the other players. We denote the amount of information available to (or estimated by) player $i$ at time $t$ with 
\begin{equation}
\bI^{(i)}_t=(I_{1,t}^{(i)}, \ldots, I_{N,t}^{(i)})\ ,\label{amount}
\end{equation}
where $I_{j,t}^{(i)}$ is the information attributed to player $j$ by player $i$ at time $t$, 
which will depend on the process outcomes up to time $t-1$ i.e. we use the short-hand notation 
$I_{j,t}^{(i)}\equiv I_j^{(i)}(\sigma_{-R,\ldots, t-1})$. 
Player $i$'s will thus estimate \eqref{eq:joint} as 
\begin{align}
& P^{(i)}(\sigma_{0,\ldots,M-1}, \bsigma_0, \ldots, \bsigma_{M-1}|\bI_0^{(i)}
)
\nonumber\\
&= P^{(i)}(\sigma_{0,\ldots, M-1}|I_i^{(i)}(\sigma_{-R,\ldots,0}))
\prod_{t=0}^{M-1} P^{(i)}(\bsigma_t|\bI_t^{(i)})\ ,
\label{eq:joint-i}
\end{align}
where the first term in the second line is the probability of the 
outcome trajectory -- as estimated by player $i$ -- on the basis 
of the information he/she has on the game, while the second term is the probability -- always estimated by player $i$ -- of the bets placed by all the players (including $i$), based on the information that player $i$ attributes to the other players and the information that player $i$ holds.

The particular form taken by \eqref{eq:joint-i} depends not only on the information held by the players, but also on the particular assumptions they make on the process, regardless of its true nature (which is generally unknown to them).
For instance, if player $i$ assumes that the process be uncorrelated in time, their estimate \eqref{eq:joint-i} will specialize to 
\begin{align}
& P^{(i)}(\sigma_{0,\ldots,M-1}, \bsigma_0, \ldots, \bsigma_{M-1}|
\bI_0^{(i)}
)\nonumber\\
& =\prod_{t=0}^{M-1} P^{(i)}(\bsigma_t|\bI_t^{(i)})P^{(i)}(\sigma_t|I_{i,t}^{(i)}
)\ , \label{eq:timeuncorrelated}
\end{align}
i.e. at each time $t$, $\sigma_t$ will be seen by player $i$ as being drawn randomly and independently from a distribution estimated from the past outcomes. 
On the other hand, if the player $i$ assumes an underlying Markovian dynamics, Eq.~\eqref{eq:joint-i} will take the form
\begin{align}
& P^{(i)}(\sigma_{0,\ldots,M-1}, \bsigma_0, \ldots, \bsigma_{M-1}|
\bI_0^{(i)})
= \nonumber\\
& \prod_{t=0}^{M-1} P^{(i)}(\bsigma_t|\bI_t^{(i)})P^{(i)}(\sigma_t|\sigma_{t-1}, I_{i,t}^{(i)}
)\ ,
\end{align}
i.e. at each time step $t$ the configuration $\sigma_t$ is seen by player $i$ as being sampled from the transition probability (from the earlier configuration $\sigma_{t-1}$), estimated from the past outcomes.
We will henceforth assume $M\ll R$ so that the data acquired from the players during the game (i.e. for $0\leq t \leq M-1$)
will not significantly affect the players' estimates, i.e. 
$P^{(i)}(\bsigma_t|\bI_t^{(i)})\simeq P^{(i)}(\bsigma_t|\bI_0^{(i)})$, 
$P^{(i)}(\sigma_t|I_{i,t}^{(i)})
\simeq P^{(i)}(\sigma_t|I_{i,0}^{(i)})$, and, similarly, $P^{(i)}(\sigma_t|\sigma_{t-1},I_{i,t}^{(i)})
\simeq P^{(i)}(\sigma_t|\sigma_{t-1}, I_{i,0}^{(i)})$. As the only temporal index appearing in the information vector will be $t=0$ to ease the notation, we will from now on drop the temporal index and write $I_{j,0}^{(i)}=I_j^{(i)}$ and $\bI_0^{(i)}=\bI^{(i)}$.

At each round, all players that have placed a successful bet are rewarded with the full pot of entry fees $N\phi$ split evenly among them. After $M$ rounds starting from time $t=0$, each player will therefore see their wealth change by the following random amount
\begin{equation}
    \Delta W_i(\sigma_{0,\ldots,M-1},\bsigma_0,\ldots,\bsigma_{M-1})= 
   -M \phi+\sum_{t=0}^{M-1} \frac{N\phi}{N_t}\delta_{\sigma_t,\sigma^{(i)}_{t}}
\ ,\label{DeltaWi}
\end{equation}
where $N_t$ is the (random) number of winners of round $t$, and $\delta_{\sigma_t,\sigma^{(i)}_{t}}=1$ if the bet placed by player $i$ at round $t$ was successful, and zero otherwise. 
The number of winners at round $t$ is given by
\begin{equation}
    N_t = \sum_{k=1}^N \delta_{\sigma_t,\sigma^{(k)}_{t}}\ .\label{eq:Ntau}
\end{equation}

The observables of interest are the average wealth changes that each player $i$ expects to realize after $M$ rounds, conditioned on the amount of information they believe every player held at each round. 
Formally
\begin{align}
\mathbb{E}_{\bI^{(i)}}^{(i)}\left[\Delta W_i\right] &= \!\!\!\sum_{
\begin{array}{c}
\sigma_{0,\ldots,M-1}\\
\bsigma_0,\ldots,\bsigma_{M-1}
\nonumber\end{array}
} 
\!\!\!\!\!\!\Delta W_i(\sigma_{0,\ldots,M-1},\bsigma_0,\ldots,\bsigma_{M-1}) \\
 & \times P^{(i)}(\sigma_{0,\ldots,M-1}, \bm\sigma_0,\ldots,\bm\sigma_{M-1} |\bI^{(i)} 
 )\ .
 \label{eq:info_expectation_wealth}
\end{align}

In this formulation of the problem, each player $i$ will only rely on information available to or estimated by them when computing their expected wealth change: the exact details of the stochastic process (unknown to all players), as well as the \emph{actual} information held by all other players do not matter.

While the discussion of our framework could be kept very general, it is now convenient to specialize the formalism to a more concrete example (a binary-outcome game) and discuss how the information exchange between players may work, and what the effect on the pricing of this information is.

\subsection{Betting on a binary-outcome game}\label{IIIa}

Assume now that, at each round $t$, the binary outcome $\sigma_t\in \{+1,-1\}$ of an underlying concealed stochastic process is announced. Prior to observing the outcome, each player $j$ will have paid the entry fee and will have declared their bet $\sigma_t^{(j)}$ on the outcome in the binary state space $\mathcal{A}=\{+1,-1\}$.

We assume that each player $i$ will regard the process as uncorrelated in time, essentially as a sequence of independent coin tosses. In addition, we assume that each player will assume the other players to bet either \textit{Head} or \textit{Tail} (respectively, $\sigma^{(i)}_t=+1$ or $\sigma^{(i)}_t=-1$) independently of each other at each round\footnote{From now on, we will talk about ``Head'', ``Tail'', and ``coin'' \emph{as if the underlying process were indeed a biased coin toss}. In reality, we stress that this is only a (natural) ``modelling'' assumption that every player may make about a stochastic process they actually know nothing about. Our formalism could accommodate different modelling choices by the players on the actual underlying process, for instance a Markov 2-state model.}. Hence, one can write the first probability appearing on the r.h.s. of  Eq.~\eqref{eq:timeuncorrelated} as
\begin{equation}
P^{(i)}(\bsigma_t|\bI^{(i)})=
\prod_{j=1}^N P_j^{(i)}(\sigma_t^{(j)}|
\bI^{(i)}
)\ ,
\label{eq:indep_bet_sequence}
\end{equation}

where $P_j^{(i)}(\sigma_t^{(j)}|
\bI^{(i)}
)$ is the probability that player $i$ believes that player $j$ will bet according to. 
This can be rewritten more explicitly in terms of  $\rho_j$, the best estimate that player $j$ will use for the bias of the coin. In particular, conditioning on $\rho_j$, we can write
\begin{align}
\nonumber P_j^{(i)}(\sigma_t^{(j)}|\bI^{(i)}
) &=
\int d\rho_j 
P_j^{(i)}(\sigma_t^{(j)}|\rho_j)P_j^{(i)}(\rho_j|\bI^{(i)})\\
&=\int d\rho_j 
P^{(i)}_{\rho_j}(\sigma_t^{(j)})f_{\bI^{(i)}}(\rho_j)\ ,
\label{eq:prob_bet_w_bias}
\end{align}
with $P^{(i)}_{\rho_j}(\sigma_t^{(j)}) \equiv P^{(i)}_j(\sigma_t^{(j)}|\rho_j)$ being the individual betting probability that player $i$ attributes to player $j$ based on his/her best estimate $\rho_j$ of the coin bias, and with $f_{\bI^{(i)}}(\rho_j)\equiv P_j^{(i)}(\rho_j|\bI^{(i)})$, the probability density function (pdf) for the optimal bias parameter $\rho_j$ that $j$ could use, 
 as estimated by player $i$ using the information he/she holds. 

 For a binary game, it is natural for every player to assume  a bimodal distribution, i.e., $
P_{\rho}^{(i)}(\sigma)\equiv\tilde{P}_\rho(\sigma)~\forall~i 
$, and that this probability will have the form
\begin{equation}
  \Tilde{P}_\varrho (\sigma) = g(\varrho)\delta_{\sigma,+1}+(1-g(\varrho))\delta_{\sigma,-1}\ ,\label{eq:strategygzero}
\end{equation}
where $g(\varrho)$ is a deterministic function of the bias parameter. It turns out (see Appendix \ref{app:strategy}) that the best choice available to players is always given by
\begin{equation}
    g(\varrho)=\theta(\varrho-1/2)\ ,\label{eq:bettingg}
\end{equation}
where $\theta(x)$ is the Heaviside step function. In other words, the best betting strategy available to players is to assume that the ``coin'' will \emph{always} come up Heads if their best estimate of the bias is $>1/2$, and Tails otherwise.

Each player $i$ will then draw their best estimate $\rho_i$, and that of their opponents $\rho_j$, of the coin bias parameter from the pdf $f_{\bI^{(i)}}(\rho_j)$. In the absence of any extra information, i.e. $\bI^{(i)}={\bm 0}$, 
it is reasonable to assume that $f_{\bm 0}^{(i)}(\rho_j)$ will be uniform in $[0,1]$ for all players and all rounds.

After $M$ rounds, each player will therefore see their wealth change by 
$\Delta W_i$, as defined in \eqref{DeltaWi}. As $\phi$ appears as an overall multiplicative constant, from now on we will set, without loss of generality, $\phi=1$. This is equivalent to setting the unit of wealth change equal to 
the fee that players pay to enter the game. 

We now assume that one of the players (say, number 1) is ``special'': they hold a long sequence of size $R\gg 1$ of past outcomes of the game, which shows a total number $H$ of Heads. Player 1 (hereafter dubbed the \emph{data holder}/\emph{seller}) has therefore two choices: (i) they can simply use this information -- without sharing it with anyone -- to improve their betting strategy by guessing a more accurate value for the coin bias, or (ii) they can trade part of this information with one colluding player (the \emph{buyer}), say player 2, in exchange for money. The additional information parameters of our game are therefore $R$ (length of the string of past outcomes of the game that player 1 holds\footnote{In the following, we are going to assume that $R\gg 1$, so by the Law of Large Numbers we expect that the data holder will always have ``good quality'' data, and therefore an edge in the game. A discussion of the ``finite horizon'' case -- where too short a string of past outcomes could actually mislead the data holder -- is deferred to a future publication.}), $r$ (length of the sub-string of data that player 1 is willing to sell to player 2), $h$ (number of Heads showing up in the string of length $r$ put up for sale), and $x$ (excess of Heads in the longer string, so $H=h+x$ 
are the Heads showing in the seller's string). 
Consequently, the information vectors ${\bf I}^{(1)}$ and ${\bf I}^{(2)}$
of players $1$ and $2$,  respectively will take values -- before the transaction\footnote{We will loosely write $I_i^{(j)}=0$ instead of the more precise $I_i^{(j)}=\emptyset$.}
\begin{align}
I_{i}^{(1)}=
    \begin{dcases}
    \{H,R\} \ &\textrm{for } i=1 \\
    0 \  &\textrm{for } i=2 \, 
    \end{dcases}
\end{align}
and
\begin{equation}
I_{i}^{(2)}=0 \quad{\rm for}~i=1,2
\end{equation}
indicating that player 2 holds no information on any player, including him/herself. After the transaction, these objects will evaluate to 
\begin{align}
I_{i}^{(1)}=
    \begin{dcases}
    \{H,R\} \ &\textrm{for } i=1 \\
    \{h,r\} \  &\textrm{for } i=2 \, 
    \end{dcases}
\end{align}
and 
\begin{align}
I_{i}^{(2)}=
    \begin{dcases}
    \{h,r\} \ &\textrm{for } i=1 \\
    \{h,r\} \  &\textrm{for } i=2
    \end{dcases}
    \ .
\end{align}
This is because the buyer is unaware of the length $R$ of the string held by the seller or the number $H$ of Heads in it, and will therefore attribute his/her own ``best'' available information ($h$ Heads showing in a string of $r$ outcomes) also to the seller.

We will use the short-hand notation for the following averages (see Eq. \eqref{eq:info_expectation_wealth})
\begin{align}
\mathbb{E}_R&=\mathbb{E}^{(1)}_{(\{H,R\},0\ldots,0)}\nonumber\\
\mathbb{E}_{R,r}&=\mathbb{E}^{(1)}_{(\{H,R\},\{h,r\}\ldots,0)}\nonumber\\
\mathbb{E}_0&=\mathbb{E}^{(2)}_{(0,\ldots,0)}\nonumber\\
\mathbb{E}_{r,r}&=\mathbb{E}^{(2)}_{(\{h,r\},\{h,r\},0,\ldots,0)}\ .\nonumber
\end{align}
We will also use a similar notation to make the dependence on the information in the expression of the coin bias pdf explicit. In particular, we define the following short-hands:
\begin{align}
    f_{R}(\rho)&= f_{(\{H,R\},0,\ldots,0)}(\rho)\nonumber\\
    f_{R,r}(\rho)&=f_{(\{H,R\},\{h,r\},\ldots,0)}(\rho)\nonumber\\
    f_{0}(\rho)&=f_{(0,\ldots,0)}(\rho)\nonumber\\
    f_{r,r}(\rho)&=f_{(\{h,r\},\{h,r\},\ldots,0)}(\rho)\ .\nonumber
\end{align}
For later convenience, we will denote with $\alpha=H/R$ the fraction of Heads in the string held by the seller. Values of $\alpha>1/2$ ($\alpha<1/2$) will signal a bias towards Head (Tail) in the seller's string.
\vspace{20pt}

\subsection{Fair price of information}
\label{sec:fair}

Our goal is now to compute the fair price at which the data holder should put up some of their data for sale, for different values of the model parameters.  
To compute this fair price, we need to determine -- from the point of view of the data holder -- the ``wealth gain'' (if any) that the data holder and the buyer may expect to realize if the transaction takes place.

The act of selling information may in principle decrease the data holder's wealth because another player (the buyer) will have a greater chance of placing a successful bet, therefore leading the fee pot to be split among a larger number of winners. This expected decrease in revenues should be (at the very least) compensated by the amount of money $\Psi_{\mathrm{min}}$ received by the buyer in exchange for the data. Similarly, the act of buying information may potentially increase the buyer's revenues, therefore there will be a maximal price $\Psi_{\mathrm{max}}$ the buyer should be fairly asked to pay for the data before their acquired edge in the game is wiped out.

The transaction should therefore be proposed by a fair seller to a prospective buyer only if $\Psi_{\mathrm{max}}\geq \Psi_{\mathrm{min}}>0$, and may happen at any price $\Psi$ such that $\Psi_{\mathrm{min}}\leq\Psi\leq \Psi_{\mathrm{max}}$.
We claim that
\begin{align}
    \Psi_{\mathrm{min}} &=\max[\mathbb{E}_{R}[\Delta W_1]-\mathbb{E}_{R,r}[\Delta W_1],0]\label{psimin}\\
    \Psi_{\mathrm{max}} &=\max[\mathbb{E}_{r,r}[\Delta W_2]-\mathbb{E}_{0}[\Delta W_2],0]\ ,\label{psimax}
\end{align}
where the max operator ensures that $\Psi_{\mathrm{min}}$, $\Psi_{\mathrm{max}}$ are non-negative (as required for prices) and 
the different expectation operators corresponding to different levels of information available to players, as explained in the earlier Section.

The meaning of Eq.~\eqref{psimin} is clear: the minimal price the data holder must impose is the average loss (if any) they will incur by selling data as opposed to holding on to it exclusively. Similarly, for Eq.~\eqref{psimax}, the maximal price that the buyer should be fairly asked to pay is the average gain (if any) they would estimate if they could use the acquired information as opposed to their original state of complete ignorance, represented by the operator $\mathbb{E}_{0}$.

We stress that the two baselines (i.e. expected wealth difference \emph{before} the transaction is arranged) for the data holder and the prospective buyer are different: while the buyer does not have any extra information about the game and would need to bet blindly unless the transaction takes place, the data holder could (and will!) still rely on the long string of past outcomes they hold exclusively to achieve a stronger betting performance. 

Even though the price curves \eqref{psimin} and \eqref{psimax} have been defined for the specific binary-outcome game, it is straightforward to generalize the reasoning to more complicated settings by comparing the expected ``wealth gains'' before and after the transaction.

\subsection{Interim summary}\label{sec:int_summary}

In summary, the class of stochastic games we are considering require specifying the following ingredients: (i) Eq. \eqref{eq:joint-i} for all players $i=1,\ldots,N$, namely the joint probability of the future outcome trajectory and the bets cast by all players at all time given the past history of the game, \emph{as perceived/estimated by each player}, (ii) The amount of information 
$\bI^{(i)}\equiv \bI^{(i)}_0$ (see Eq. \eqref{amount}) available to each player about all the others, prior to the start of the game, 
and (iii) the rule by which two (or more) players may exchange part of their exclusive information, namely how the information vectors would change between before and after the transaction. Given these ingredients, it is possible to compute the $\Psi_{\mathrm{min}}$ and $\Psi_{\mathrm{max}}$ price curves by comparing the expected wealth changes perceived by each player involved in the transaction. For the binary-outcome game described in more detail here, these curves are determined by Eqs. \eqref{psimin} and \eqref{psimax} and will be computed explicitly below.

In the next section, we summarize our results on the limiting price curves, for the case of a single round $M=1$ of the binary-outcome game described in detail above. Specifically, we will discuss different choices of the values of the parameters and the regions where the transaction may or may not take place.
 In the following sections, we will compute all the averages appearing in Eqs.~\eqref{psimin} and \eqref{psimax} explicitly.

\section{Summary of results}\label{sec:summary}

Consider a single round ($M=1$) of the game with $N=5$ players. Player 1 holds a string of $R=100$ past outcomes of the coin toss, which shows $H=h+40$ Heads. Here, $h\in\{0,1,\ldots,60\}$ is the number of Heads showing in the sub-string of length $r=60$ being potentially traded with the buyer (Player 2). 

\begin{figure}[h]
    \centering
    \includegraphics[scale = 0.3]{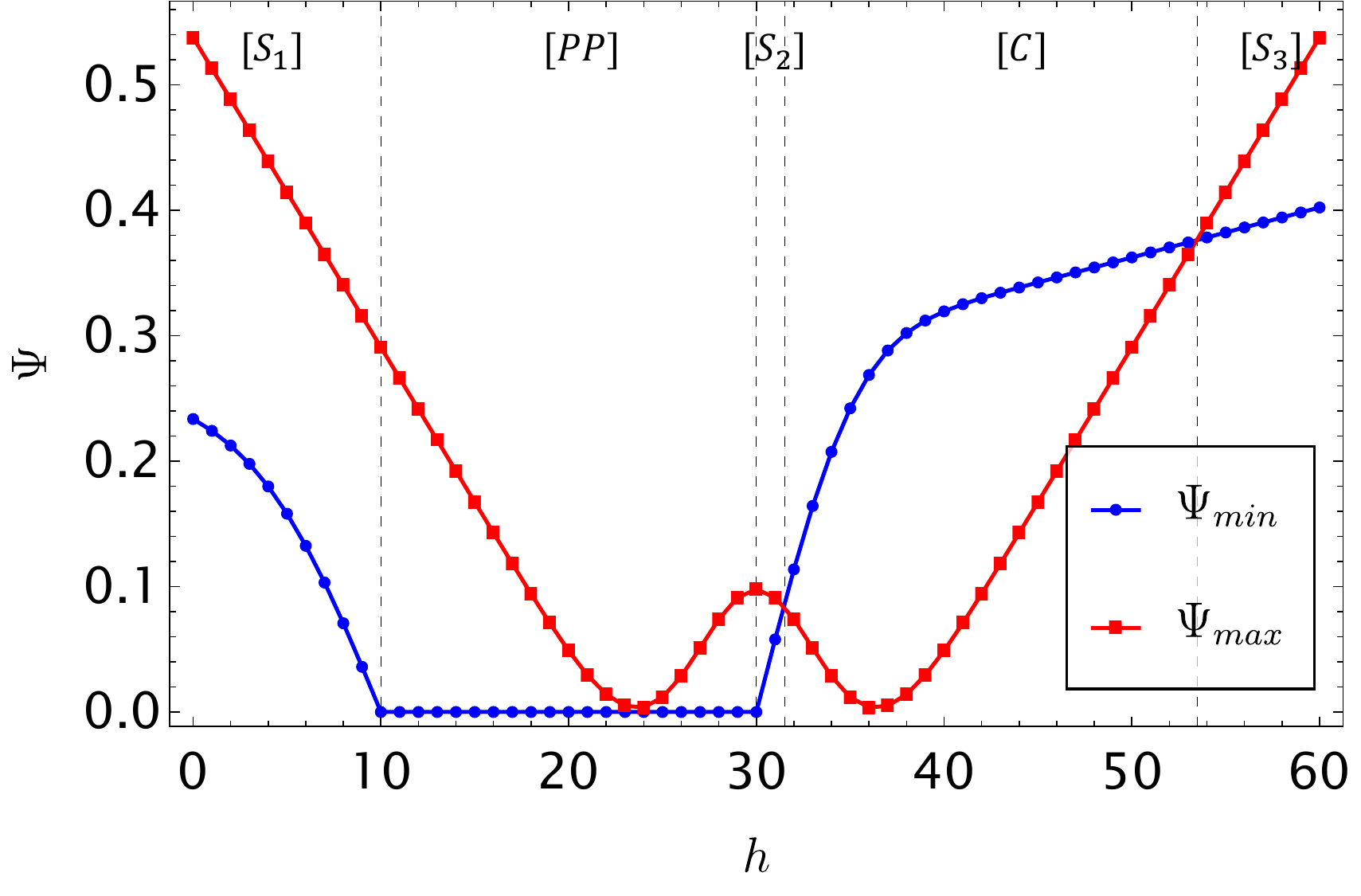}
    \caption{Limiting price curves $\Psi_{\mathrm{min}}$ (blue) and $\Psi_{\mathrm{max}}$ (red) from Eqs.~\eqref{psimin} and \eqref{psimax} for $M=1$, $\phi=1$, and $N=5$ as a function of $h$, the number of Heads showing in the sub-string of length $r=60$ offered for sale. The number of extra Heads, unobserved by the buyer, is fixed to $x=40$ in a  string of length $R=100$, which is owned by the seller. The explicit expressions for the curves are reported in Appendix \ref{app:LimitingPrice}. }
    \label{fig:pricevsh}
\end{figure}

In Fig. \ref{fig:pricevsh}, we show the limiting price curves $ \Psi_{\mathrm{min}} $ (blue) and $ \Psi_{\mathrm{max}} $ (red) in Eq.~\eqref{psimin} and \eqref{psimax}, as a function of $h$ and for $\phi=1$. The analytical expressions of $ \Psi_{\mathrm{min}} $ and $ \Psi_{\mathrm{max}} $ as functions of the parameters of the game are reported in Appendix \ref{app:LimitingPrice}. 
We observe five different regimes:
\begin{itemize}
    \item {\bf Regime [$S_1$] - Symbiosis:} The transaction should take place, as it will be profitable for both players. In this regime, both the buyer and the seller would agree that the coin must be strongly Tail-biased -- as the buyer observes at most $10$ Heads in a string of $r=60$ past outcomes, and the seller observes at most $50$ Heads in a string of $R=100$ past outcomes. This information assists both players in placing a favorable bet and the buyer upside opportunity is enough to cover the loss of edge held by the seller should the transaction not occur. As a result, any price point $\Psi$ between the blue (seller) and red (buyer) curves in Fig. \ref{fig:pricevsh} would realize a favorable transaction for both parties.
    \item {\bf Regime [$PP$] - ``Prey-Predator'':} This is a region where the substring that the data holder should put up for sale displays the \emph{opposite} bias to the original (longer) one that they own. 
    In this situation, an exploitative seller (predator) could, in principle, sell the string at a positive price or even give it away for free, realizing a profit by tricking a  potential competitor (the buyer) into betting on an unlikely outcome. Indeed, had the buyer access (post transaction) to the seller's knowledge, they would ascertain that the absolute best strategy in this regime would have been to refuse the transaction altogether. This is a classical \emph{caveat emptor} situation, where the purchase of a defective good -- genuine (non-doctored) yet misleading data -- is the buyer's responsibility unless the seller had made an active misrepresentation of the data quality. 
    This situation would not occur, however, under our assumption that the seller is fair and reputable. Realizing the information they hold is at best worthless for the buyer, if not outright misleading, they would not put their substring up for sale in this regime. 
    \item {\bf Regime [$S_2$] - Symbiosis:} The transaction should take place again. In this regime, both the buyer and the seller are in agreement that the coin must be Head-biased and they will both profit from exchanging this information for a fair price.
    \item {\bf Regime [$C$] - Competition:} This is a regime of high competition, in which the data holder would need to receive a large sum of money to cover the potential loss of edge. From the buyer's perspective, the upside opportunity in this region is not enough to justify the investment in the data and thus the transaction would not occur. In fact, both the buyer and the seller are in agreement that the coin must be Head-biased and both would bet on the same outcome, creating a ``destructive interference'' between these two special players. 
    \item {\bf Regime [$S_3$] - Symbiosis:} The transaction would take place. The situation is mirror-symmetric with respect to Regime [$S_1$], the Head bias of the coin being so strong and apparent that, even though the seller has great confidence in the final outcome of the game, the buyer would still be offering an amount of money  that would cover for the seller's lowered expected return.
\end{itemize}

\begin{figure}[h]
    \centering
    \includegraphics[scale = 0.3]{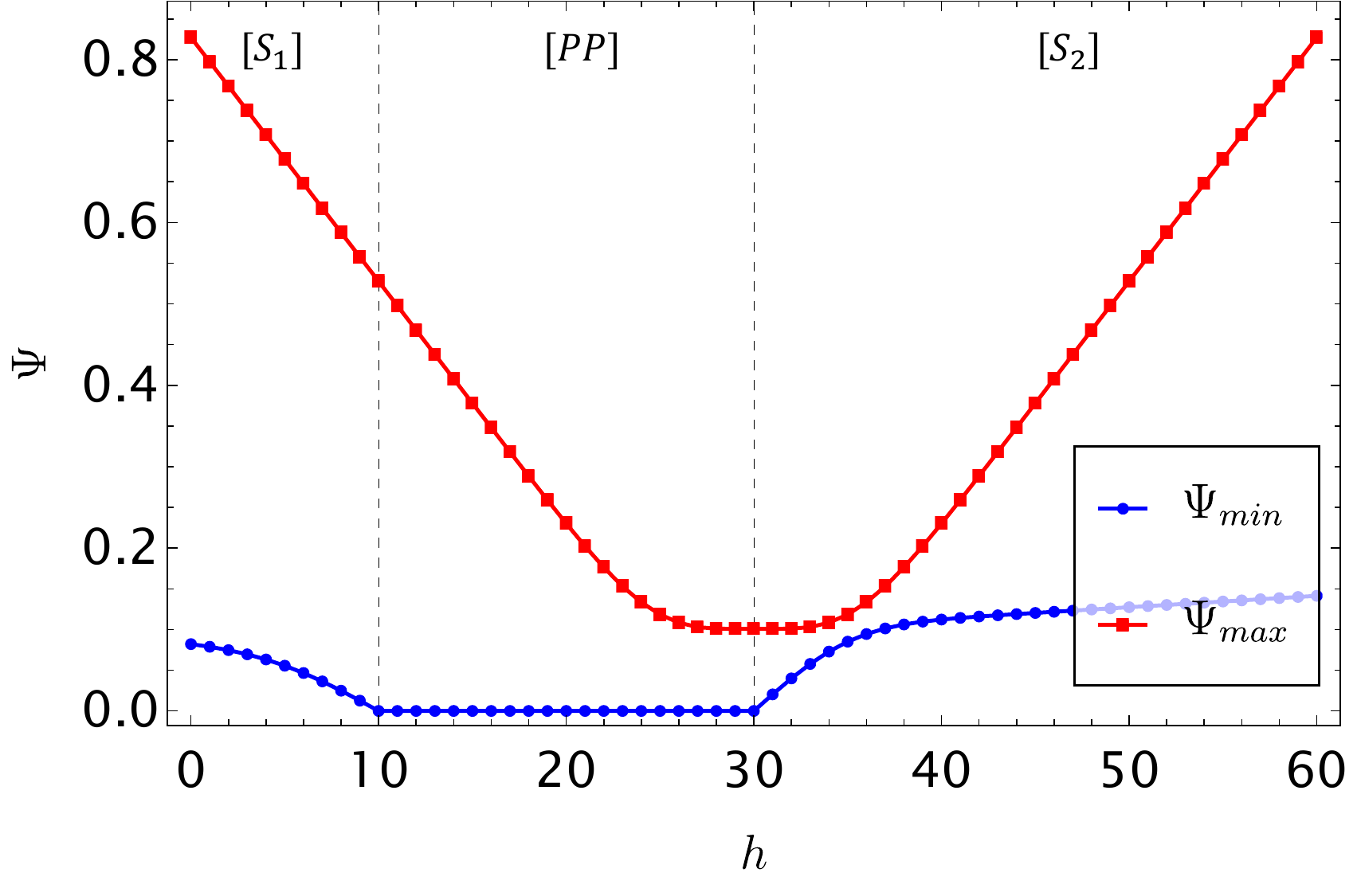}
    \caption{ Limiting price curves $\Psi_{\mathrm{min}}$ (blue) and $\Psi_{\mathrm{max}}$ (red) from Eqs.~\eqref{psimin} and \eqref{psimax} for $M=1$, $\phi=1$, and $N=15$ as a function of $h$, the number of Heads showing in the sub-string of length $r=60$ offered for sale.  The number of extra Heads, unobserved by the buyer, is fixed to $x=40$ in a  string of length $R=100$, which is owned by the seller. The explicit expressions for the curves are reported in Appendix \ref{app:LimitingPrice}. }
    \label{fig:pricevsh2}
\end{figure}

We note a surprising non-monotonic behavior of $\Psi_{\mathrm{max}}$ around $h=r/2$: seemingly, the buyer is inclined to pay more for information that is completely unbiased than for information that is {\it slightly} biased. The interpretation is that any slight bias will induce the buyer to think that the seller 
will be playing the same strategy as him/her, and therefore the buyer will perceive a higher competition with the seller, which is not balanced by a sufficiently high expectation to win the bet. Upon increasing the number of players $N$, such direct competition (for finite resources) between the buyer and the seller becomes weaker, as the pool of resources increases with $N$. Numerical analysis shows that indeed for $N$ above a certain threshold, $\bar{N}$, $\Psi_{\mathrm{max}}$ transitions from a double well to a single well function.

Consider now a single round ($M=1$) but with a larger number of players ($N=15$), still keeping the same parameters as before ($R=100$, $r=60$, and $x=40$). In this case (see Fig. \ref{fig:pricevsh2}), there are no longer situations where the transaction is unappealing to the data holder -- as was the case in Regime [$C$] beforehand. We still observe two ``symbiotic'' regimes ([$S_1$] and [$S_2$]) and a ``prey-predator'' regime, with the same features as before.

\begin{figure}[h]
    \centering
    \includegraphics[scale = 0.45]{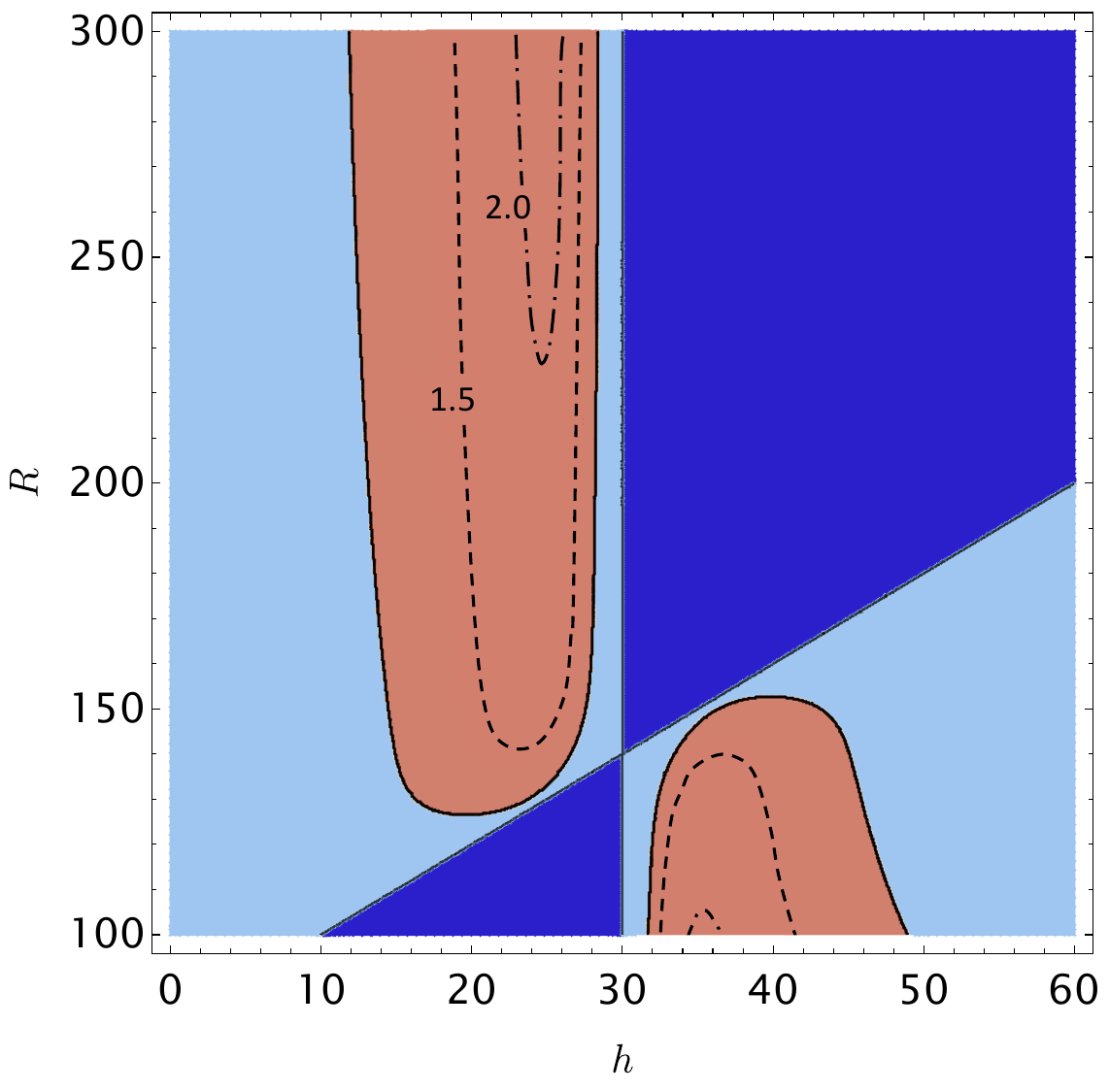}
    \caption{Phase diagram in the parameter space $(h, R)$, 
    where $R$ is the length of the string held by the seller and $h$ is the number of Heads in the buyer's string. The red-shaded area, delimited by the line $N^\star/N=1$ denotes the 
    phase where destructive competition emerges. Contours within this region indicate the lines $N^\star/N=1.5$ and $N^\star/N=2$. Outside the red shaded area, the game is in the phase where it is always appealing to the data holder. The dark blue region corresponds to the ``prey-predator'' regime, while the light blue region indicates  the ``symbiotic" regime.
    The parameters are set as follows: the number of players is $N=6$, the number of extra Heads in the seller's string (unobserved by the buyer) is $x=40$ and the length of the string put on sale is $r=60$. 
    }
    \label{fig:Nstarpanel}
\end{figure}

The comparison between Fig. \ref{fig:pricevsh} and \ref{fig:pricevsh2} shows that a sharp \emph{phase transition} may occur in the number $N$ of players, such that for $N<N^\star$ we expect a ``competition'' regime to materialize, which hinders the prospective transaction, whereas for $N\geq N^\star$ the transaction is always appealing to the data holder, and -- only in the ``symbiotic'' regimes -- also favorable to the buyer. The fact that a more crowded game may remove the ``destructive competition'' regime between the two special players has an interesting interpretation in the context of the ecology of the game, as it simply signals that more 
unsuspecting players become available to prey upon. The amount of available resources in a ``smaller'' game -- namely, the fees brought in by the $N-2$ uninformed players -- may be insufficient to accommodate a symbiotic/cooperative interaction between the two special players, who are instead better off relying on their own means. For the same reason, the data holder is willing to sell at a lower price the higher the number of players. As $\Psi$ in Fig. \ref{fig:pricevsh2} is double-well (with maximum and minima seen to lie almost on a horizontal line, on this scale) we have that $N^\star<\bar{N}$, i.e. destructive competition is removed at a lower value of $N$ than the value $\bar{N}$ at which $\Psi$ becomes single-well.

In Fig. \ref{fig:Nstarpanel}, we further investigate the seller-buyer interaction and the behavior of the critical value $N^\star$ as a function of $R$ (the length of the string held by the data holder) and $h$ (the number of Heads in the buyer's string), 
for a fixed value of $N$ (the number of players). For $N^\star\geq N$ (red shaded area) the game is in the phase where destructive competition emerges. The contour lines $N^\star/N=1.5$ and $N^\star/N=2$ show that $N^\star$ grows as one moves deeper inside this region. Hence, for fixed $h$, $N^\star$ is a monotonic function of $R$ which increases when $h< r/2$ and decreases for $h>r/2$, whereas for fixed $R$, $N^\star$ is a non-monotonic function of $h$. Outside of the red-shaded area, 
the game is always appealing and, in particular, is characterized by the prey-predator regime in the blue region and the symbiotic regime in the light blue region. Interestingly, this 
occurs both at large and small values of the bias in the string that is put on sale (i.e. $h$ far away from $r/2$ and close to it, respectively). 

As a final remark, we note that the two curves shown in Figs. \ref{fig:pricevsh} and \ref{fig:pricevsh2} can only be known to the seller. The buyer could only calculate the family of curves $\Psi_{\rm max}$ for different values of $h$ and $r$ (which are however unknown to him/her until the transaction occurs). This is however not a limitation in the context of our setting, where it is assumed that a reputable seller proposes a fair price to a prospective buyer (see Sec. \ref{sec:fair}).

In the next section, we compute explicitly the two baseline pay-offs for the prospective seller and buyer that appear in Eqs. \eqref{psimin} and \eqref{psimax}.

\section{Calculation of Baselines}\label{sec:baseline}

\subsection{The prospective buyer}

We start by computing the baseline $\mathbb{E}_{0}[\Delta W_2]$ appearing in Eq.~\eqref{psimax} from Eq.~\eqref{eq:info_expectation_wealth}. This is the wealth difference (gain/loss) that the prospective buyer (or any other uninformed player) expects to make placing a bet, in the absence of any information about the previous history of the game or the other players, and without any transaction being arranged.

By linearity of expectations, this requires computing 
$   \mathbb{E}_{0}\left[\frac{\delta_{\sigma_t,\sigma^{(2)}_{t}}}{N_t}\right],
$
where $N_t$, the number of winners at round $t$ is given in Eq.~\eqref{eq:Ntau}. The expectation needs to be computed over the full joint distribution for the (uncorrelated) outcomes and the bets $P^{(i)}(\sigma_{0,\ldots,M-1}, \bsigma_0, \ldots, \bsigma_{M-1}|\bI^{(i)})$ for $i=2$, introduced in Eq.~\eqref{eq:timeuncorrelated}
equipped with  Eq.~\eqref{eq:indep_bet_sequence}
and Eq.~\eqref{eq:prob_bet_w_bias}
where $\bI^{(2)}=\bm 0$. 
Given the factorization over time in Eq.~\eqref{eq:timeuncorrelated}, such expectation boils down to computing the expectation over the marginal distribution at time $t$,
$P^{(i)}(\bsigma_t|\bI^{(i)})P^{(i)}(\sigma_t|I_{i}^{(i)})$.

With our assumptions, player $2$ would not have any other choice but to use the \emph{same} estimate $\tilde P_{\rho_2}$ from \eqref{eq:strategygzero} for both his/her betting strategy, \emph{and} the actual outcome of the process. Ordinarily, it would not make sense to bet on a process according to a probability law that is different from the best estimate one has of how the process actually unfolds!

The only other possibility would be for player $2$ to assume that the underlying process is actually an independent (biased) coin toss happening with probability 
\begin{equation}
    P_\rho(\sigma_t)=\rho \delta_{\sigma_t,+1}+(1-\rho)\delta_{\sigma_t,-1}\ .
\end{equation}
The obvious problem with this choice would be that the expected wealth changes are going to depend on the actual bias $\rho$ of the ``coin'', which no player actually knows. However, it turns out that -- for the specific calculation of the baseline below -- the actual coin bias $\rho$ drops out of the final expression, therefore this second (ordinarily precluded) avenue becomes actually viable for player $2$.

Let us therefore write the expectation explicitly, with the newly introduced notation for the actual outcome probability, without replacing it with any proxy

\begin{widetext}
\begin{align}
\mathbb{E}_{0}\left[\frac{\delta_{\sigma_t,\sigma^{(2)}_{t}}}{N_t}\right] &=\int_0^1 d\rho_1~f_0(\rho_1)\cdots\int_0^1 d\rho_N~f_0(\rho_N)\sum_{\sigma_t=\pm 1}P_\rho(\sigma_t)\sum_{\bm\sigma_t\in \{\pm 1\}^N}\tilde P_{\rho_1}(\sigma_t^{(1)})\cdots \tilde P_{\rho_N}(\sigma_t^{(N)})\frac{\delta_{\sigma_t,\sigma^{(2)}_{t}}}{\sum_{k=1}^N \delta_{\sigma_t,\sigma^{(k)}_{t}}}\ .
\label{eq:expzerorho1}
\end{align}
\end{widetext}
For uniform pdfs $f_0(\varrho)$ of the estimates of the coin bias made by all players -- the $\rho$-dependence completely drops out\footnote{If this fortunate dropping-out did not happen, we should have performed the calculation \emph{from the point of view of player 2} -- dubbed $\mathbb{E}_{0,\cancel{\rho}}[\cdots]$ -- who would have used their own estimate $\rho_2$ for the coin bias in lieu of the actual (unknown) coin bias $\rho$. The calculation of this case is performed in Appendix \ref{app:E0} as well, and leads to a higher pre-factor ($3/4$ instead of $1/2$) in Eq.~\eqref{eq:E0right}. It is therefore natural to assume that player $2$ will use his/her more ``conservative'' estimate in Eq.~\eqref{eq:E0right} as opposed to the more generous estimate in \eqref{eq:upside_lucky}, as the latter would be inevitably inflated by the use of the same distribution for one's own betting strategy and the actual outcome of the process.}, and we eventually find (see Appendix \ref{app:E0} for details)

\begin{equation}
\mathbb{E}_{0}\left[\frac{\delta_{\sigma_t,\sigma^{(2)}_{t}}}{N_t}\right] =\frac{1}{2}\frac{2-2^{1-N}}{N}\ ,\label{eq:E0right}
\end{equation}
from which it follows that the baseline for player 2 reads
\begin{equation}
\mathbb{E}_{0}\left[\Delta W_2\right] = -M\frac{1}{2^N}\ .
    \label{eq:true_baseline}
\end{equation}

In the next section, we will compute the baseline pay-off for the only player that owns some information about past outcomes.

\subsection{The prospective seller (data holder)}

In this section, we compute the baseline $\mathbb{E}_{R}[\Delta W_1]$ appearing in Eq.~\eqref{psimin} from Eq.~\eqref{eq:info_expectation_wealth}. This is the wealth gain that player 1 -- the data holder -- expects to achieve by exploiting exclusively the time series of the past $R$ outcomes of the coin toss $\sigma_{-R, \ldots,-1}$. 
Given that the data holder has recorded exactly $H$ Heads out of the past $R$ outcomes ($H\in\{0,1, \ldots,R\}$), they will build an estimate of the coin bias pdf $f_R(\rho_1)$ based on the number of Heads $H$ observed in the string.
Using Bayes' theorem
\begin{equation}
f_R(\rho_1) = \frac{P(H|R,\rho_1) f(\rho_1)}{\int_0^1 P(H|R,\varrho) f(\varrho)d\varrho}\ ,
    \label{eq:seller_posterior0}
\end{equation}
where the probability $P(H|R,\varrho)$ of obtaining $H$ heads in $R$ tosses of a coin with bias $\varrho$ is given by  
 \begin{equation}
    P(H|R,\varrho) = \binom{R}{H} \varrho^H (1 - \varrho)^{R-H}\ ,
    \label{eq:binomial}
\end{equation}
while $f(\varrho)$ is the prior probability density function for the coin bias, which we may assume uniform.
Hence, the posterior follows as
\begin{align}
    f_R(\rho_1) &= \frac{\rho_1^H (1 - \rho_1)^{R-H}}{\int_0^1\varrho^H (1 - \varrho)^{R-H} d\varrho} \nonumber\\
    &=  \frac{(R+1)!}{H!(R-H)!}\rho_1^H (1-\rho_1)^{R-H}\ ,
    \label{eq:seller_posterior}
\end{align}
which is correctly normalized, $\int_0^1 d\rho_1 f_R(\rho_1)=1$.

Similarly to the previous case, the expectation requires computing
$   \mathbb{E}_{R}\left[\frac{\delta_{\sigma_t,\sigma^{(1)}_{t}}}{N_t}\right],
$
in analogy with Eq.~\eqref{eq:expzerorho1} but with $f_0(\rho_1)$ replaced by the data holder's best estimate of the actual coin bias pdf.

Although this calculation is possible, it yields a result that this time depends explicitly on the actual coin bias $\rho$, at odds with what happened for the ``zero-information'' case (see Eq.~\eqref{eq:E0right} and Appendix \ref{app:E0}). Since no player knows the actual bias of the coin, nobody can estimate their expected gain using this piece of information \emph{unless $\rho$ drops out of the final expression}: as pointed out in the previous paragraph, we are therefore forced to assume that the data holder will use their best estimate $\rho_1$ for the coin bias in both their own betting strategy and as a proxy for the true coin bias, which will in general lead to an over-estimate of their winning chance. 

To compute this expectation, we can start from Eq.~\eqref{eq:expzerorho1} with $\sigma_t^{(2)}$ replaced by $\sigma_t^{(1)}$ to indicate the focus on the seller's bet, the uniform (``zero information'') pdf $f_0(\rho_1)$ replaced by the posterior $f_R(\rho_1)$ in Eq.~\eqref{eq:seller_posterior}, and $P_\rho(\sigma_t)$ replaced by $P_{\rho_1}(\sigma_t)$. Eventually, we obtain
\begin{equation}  \mathbb{E}_{R}\left[\frac{\delta_{\sigma_t,\sigma^{(1)}_{t}}}{N_t}\right]
    =\frac{2-2^{1-N}}{N} \Xi_R(H) \ ,
\label{eq:ERseller}
\end{equation}
\begin{figure}[h]
    \centering
    \includegraphics[scale = 0.6]{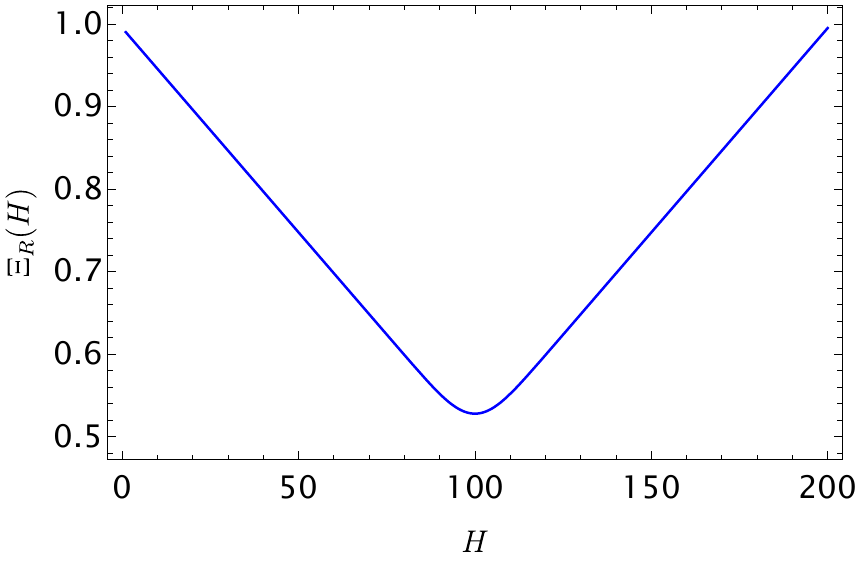}
    \caption{Expected win rate from Eq.~\eqref{XiRHtext} as a function of the number of Heads $H$, in the string of length $R=200$ held by the seller (data holder).
    }
    \label{fig:expwinXiR}
\end{figure}
where
\begin{align}    \nonumber &\Xi_R(H)=\mathbb{E}_R[\delta_{\sigma_t,\sigma_t^{(1)}}]=\\
&=\int_0^1d\rho_1 f_R(\rho_1)\left[\rho_1 g(\rho_1) + (1-\rho_1)(1-g(\rho_1))\right]
\label{eq:seller_winning_prob}
\end{align}
is the winning probability estimated by the data holder (without any knowledge of the actual coin bias) and $g(\varrho)$ is the betting parameter defined in Eq.~\eqref{eq:bettingg} -- see Appendix \ref{app:winning_probability} for more detail. 
This is given by
\begin{align}
\nonumber  \Xi_R(H) &=\frac{(R+1)!}{H! (R-H)! 2^{R+2}}\left[F(H+1,R-H)+\right.\\
  &\left. +F(R-H+1,H)\right]\ ,\label{XiRHtext}
\end{align}
where
\begin{equation} 
F(x,y)=\int_0^1 dt~(1+t)^x (1-t)^y=\frac{\, _2F_1(1,-x;y+2;-1)}{y+1}\ ,
  \label{eq:F_definition}
 \end{equation}
in terms of a hypergeometric function defined as
\begin{equation}
    _2F_1(a_1,a_2;b_1;z)=\sum_{\kappa=0}^\infty \frac{(a_1)_\kappa(a_2)_\kappa}{(b_1)_\kappa}\frac{z^\kappa}{\kappa!}\ ,
\end{equation}
where $(a)_n$ is the $n$-th order Pochhammer polynomial. 

In Fig. \ref{fig:expwinXiR}, 
we plot the expected win probability of the data holder, as given in \eqref{XiRHtext}, for a string of length $R=200$. 
The plot is consistent with the limiting behaviour of $\Xi_R(H)$ for large $R$, which can be calculated as (see Appendix \ref{app:winning_probability} for details) 
\begin{equation}
    \Xi_{R}(H=\alpha R)\sim \alpha\theta(\alpha-1/2)+(1-\alpha)\theta(1/2-\alpha)\ .\label{XiRasympt}
\end{equation}
This behavior is in line with our expectations: in the extreme cases $\alpha=0$ and $\alpha=1$ (no Head, or no Tail observed in a long string of size $R$), the data holder will rightfully conclude that the coin will always come up Tail or Head, respectively, and by betting accordingly they expect to win every time $(\Xi_{R}(H=0,1))\to 1$. Increasing $\alpha$ from zero (or decreasing $\alpha$ from $1$, symmetrically) signals a coin that is strongly biased, but will not come up Heads (or Tails) every single time: this leads the data holder's winning expectation to deteriorate linearly, up until the maximally uncertain situation $\alpha=1/2$, where the winning probability is $50-50$. Interestingly, for $\alpha=0.5$ (when there is no bias/information to be exploited in the seller's data) and \emph{finite} $R$, the expected win rate is still higher than $1/2$.  

Inserting Eq.~\eqref{eq:ERseller} into Eq.~\eqref{DeltaWi}, we can now compute the expected wealth difference for the data holder after $M$ rounds from their own viewpoint
\begin{equation}
    \mathbb{E}_R\left[\Delta W_1 \right] = M (2-2^{1-N})\Xi_{R}(H) -M\ .
\end{equation}
We remind that we have assumed that the estimates be made before playing the first of the $M$ rounds of the game and not be updated as the game progresses --- thus using only the information that is initially available to the data holder.

In the next section, we are going to compute the data holder's estimates of their own expected wealth difference -- as well as the buyer's -- should their proposed transaction go through, which will feed into the limiting prices in Eqs.~\eqref{psimin} and \eqref{psimax}.

\section{Expected wealth changes after the transaction is arranged}\label{sec:expectedafter}

\begin{figure}[h]
    \centering
    \includegraphics[scale = 0.6]{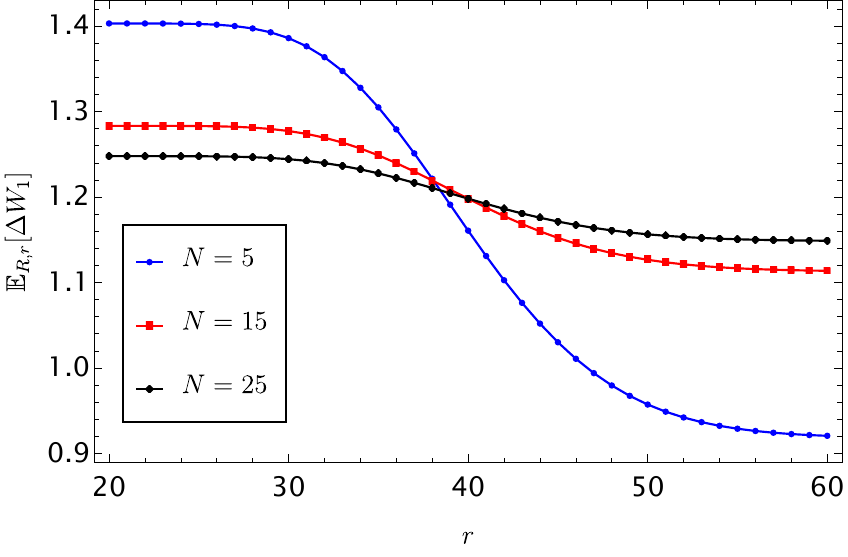}
    \caption{Expected wealth difference $\mathbb{E}_{R,r}[\Delta W_1]$ as a function of the length $r$ of the string put up for sale, and fixed number of Heads, $h=20$, for different values of $N$, as shown in the legend. The seller holds a string with total length $R=200$ and a total number of Heads $H=80$. 
    }
    \label{fig:ERr}
\end{figure}

We now imagine that a deal is proposed by 
the data holder (hereafter, the \emph{seller}) to player 2 (hereafter, the \emph{buyer}), whereby some portion of the data held by the seller is shared with the buyer for a fee. In particular, the seller would put up a sub-string of $r\leq R$ elements of the recorded outcomes $ \sigma_{-r, \ldots, -1}$ for sale. In our specific setting, it is not essential that the outcomes put up for sale be sequential. The seller could as well \emph{randomly} select $r$ elements from the string of outcomes they hold. However, in more complex games (e.g. Markov chains, processes with memory...) and for more complex betting strategies (e.g. Markov State Models), the exact sequence and ordering of data may become relevant. Note however that in our general setting the seller does not ever have the power to cherry-pick or engineer the entries of the sub-string put up for sale.

We now wish to compute the estimates made by the seller of their own expected wealth change -- as well as the buyer's -- if the transaction went through.

We now rewrite the posterior in Eq.~\eqref{eq:seller_posterior0} minding the information that the holder exchanges with the buyer
 \begin{equation}
    P(x,h|R,r,\varrho) = \binom{r}{h}\varrho^h (1 - \varrho)^{r-h}\binom{R-r}{x} \varrho^{x} (1 - \varrho)^{R-r-x}\ ,
    \label{eq:binomial_joint}
\end{equation}
with $x+h=H$. This is the probability that a string of coin outcomes of length $R$ includes a sub-string of length $r$, with $h$ heads in the sub-string of length $r$ and $x=H-h$ heads in the remaining sub-string of length $R-r$, if the tosses are generated with bias $\varrho$. Note that the distribution in Eq.~\eqref{eq:binomial_joint} is correctly normalized as
\begin{align}  
\nonumber &\sum_{h=0}^r\sum_{x=0}^{R-r} P(x,h|R,r,\varrho) =\sum_{h=0}^r\binom{r}{h}\varrho^h (1 - \varrho)^{r-h}\times\\
\nonumber &\times\sum_{x=0}^{R-r}\binom{R-r}{x} \varrho^{x} (1 - \varrho)^{R-r-x}=1\ .
\end{align}

This can be inverted using Bayes' theorem, and -- assuming the priors to be uniform -- we get, similarly to Eq.~\eqref{eq:seller_posterior},
\begin{align}
    f_{R,r} &(\rho_1) = \frac{\binom{r}{h}\binom{R-r}{x} \rho_1^{h+x} (1 - \rho_1)^{R-(h+x)}}{\int_0^1\binom{r}{h}\binom{R-r}{x} \varrho^{h+x} (1 - \varrho)^{R-(h+x)} d\varrho} \nonumber\\
    &=  \frac{(R+1)!}{(h+x)!(R-(h+x))!}\rho_1^{h+x} (1-\rho_1)^{R-(h+x)}\ .
    \label{eq:bias_distr_Rr}
\end{align}
This is the pdf of the coin bias as estimated by the seller, given the information available to them.

On the other hand, the buyer would estimate the coin bias as being drawn by the posterior in Eq.~\eqref{eq:seller_posterior} with $R,H$ replaced by $r,h$, namely
\begin{equation}
    f_{r,r}(\rho_2)=\frac{(r+1)!}{h!(r-h)!}\rho_2^h (1-\rho_2)^{r-h}\ .
    \label{eq:bias_distr_rr}
\end{equation}

\subsubsection{Seller's wealth change}

We firstly compute the seller's estimate of their wealth change $\mathbb{E}_{R,r}[\Delta W_1]$, appearing in \eqref{psimin}, assuming that the transaction would take place. This can be computed from Eq.~\eqref{eq:info_expectation_wealth}, but in this case using the bias posterior of Eq.~\eqref{eq:bias_distr_Rr} for the seller own estimate and the buyer bias posterior introduced in Eq.~\eqref{eq:bias_distr_rr} to inform the buyer's betting strategy. 
In formulae, we proceed as usual, by using the identity $\frac{1}{x} = \int_0^\infty ds~ e^{-sx}$ for $x>0$ to lift up the denominator $N_t$, then compute the expectation
\begin{widetext}
\begin{align}
&\nonumber\mathbb{E}_{R,r}\left[\frac{\delta_{\sigma_t,\sigma^{(1)}_{t}}}{N_t}\right] =\int_0^\infty ds \int_0^1 d\rho_1~f_{R,r}(\rho_1)\int_0^1 d\rho_2~f_{r,r}(\rho_2)\cdots\int_0^1 d\rho_N~f_0(\rho_N)\sum_{\sigma_t=\pm 1} P_{\rho_1}(\sigma_t)\sum_{\sigma_t^{(1)}=\pm 1}\tilde P_{\rho_1}(\sigma_t^{(1)})\times\\
&\times \delta_{\sigma_t,\sigma_t^{(1)}}e^{-s \delta_{\sigma_t,\sigma_t^{(1)}}} \sum_{\sigma_t^{(2)}} \tilde P_{\rho_2}(\sigma_t^{(2)})e^{-s \delta_{\sigma_t,\sigma_t^{(2)}}}\cdots \sum_{\sigma_t^{(N)}} \tilde P_{\rho_N}(\sigma_t^{(N)})e^{-s \delta_{\sigma_t,\sigma_t^{(N)}}}\ .
\label{eq:ERrtext}
\end{align}
\end{widetext}

The calculation is performed in Appendix \ref{app:xy_seller_exp} and gives
\begin{equation}   \mathbb{E}_{R,r}\left[\frac{\delta_{\sigma_t,\sigma^{(1)}_{t}}}{N_t}\right]=C_N X_{R,r}(x,h)+D_N Y_{R,r}(x,h)\ ,
\label{eq:ERr}
\end{equation}
with the coefficients $C_N$ and $D_N$ defined in Eqs.~\eqref{eq:Cn} and \eqref{eq:Dn}, and the final expressions for $X_{R,r}$ and $Y_{R,r}$ in terms of hypergeometric functions given in Eqs.~\eqref{eq:XRrapp} and \eqref{eq:YRrapp} respectively.

The expected wealth difference is higher in correspondence of the ``prey-predator'' region, as the buyer 
faces a more significant risk of losing in the competition for the prize when sold deceptive information. However, this impact diminishes as the number of players increases, leading to a flattening of the seller's wealth curve.

\subsubsection{Buyer's wealth change}
We still need to determine the expected change in wealth for the buyer, assuming the transaction proceeds, denoted as $\mathbb{E}_{r,r}[\Delta W_2]$, as appearing in Eq.~\eqref{psimax}. This expectation can be interpreted from the two equivalent viewpoints (i) the \emph{ex-post} buyer's perspective, i.e. as if the buyer could actually look into the sub-string being put up for sale and use the only information therein to estimate their own strategy, the seller's, and the actual coin bias, therefore being retroactively able to assign a monetary value to the quality of data they received, or -- more realistically -- (ii) the \emph{ex-ante} seller's viewpoint, i.e. assuming that the seller puts themselves in the (future) buyer's shoes and aims to know what the buyer should be fairly asked to pay for the quality of information they will find in the $r$ sub-string when they receive it.  As previously remarked, there exists an 
\emph{informational asymmetry} between the seller and the buyer, and we refrain from endorsing any specific 
viewpoint 
or disclosure rule. 
The framework developed here is intended to be 
general and it aims at 
determining the potential price range at which information transactions can, in principle, be advantageous for either biased player. In fact, both parties can independently construct their price curves based on the number of Heads in the sub-string offered for sale, treating it as a free parameter, even in cases where a sale is not ultimately agreed upon.

A key consideration in computing this final expectation involves selecting the appropriate bias posterior pdf for the seller. We have noted that post-transaction, the buyer would still lack access to the full scope of the seller's information and cannot ascertain the amount of information the seller possesses. Nevertheless, it is prudent for the buyer to operate under the assumption that the seller will engage with -- at the very least -- the same level of knowledge to shape their strategy. Consequently, the bias posterior pdfs that should be employed for both the seller and the buyer in these calculations are specified by Eq.~\eqref{eq:bias_distr_rr}. The buyer's expected wealth change is thus determined by the expectation
\begin{widetext}
\begin{align}
&\nonumber\mathbb{E}_{r,r}\left[\frac{\delta_{\sigma_t,\sigma^{(2)}_{t}}}{N_t}\right] =\int_0^\infty ds \int_0^1 d\rho_1~f_{r,r}(\rho_1)\int_0^1 d\rho_2~f_{r,r}(\rho_2)\cdots\int_0^1 d\rho_N~f_0(\rho_N)\sum_{\sigma_t=\pm 1} P_{\rho_2}(\sigma_t)\sum_{\sigma_t^{(2)}=\pm 1}\tilde P_{\rho_2}(\sigma_t^{(2)})\times\\
&\times \delta_{\sigma_t,\sigma_t^{(2)}}e^{-s \delta_{\sigma_t,\sigma_t^{(2)}}} \sum_{\sigma_t^{(1)}} \tilde P_{\rho_1}(\sigma_t^{(1)})e^{-s \delta_{\sigma_t,\sigma_t^{(1)}}}\cdots \sum_{\sigma_t^{(N)}} \tilde P_{\rho_N}(\sigma_t^{(N)})e^{-s \delta_{\sigma_t,\sigma_t^{(N)}}}\ .\label{eq:ErrWealthBuyer}
\end{align}
\end{widetext}

\begin{figure}[h]
    \centering
    \includegraphics[scale = 0.6]{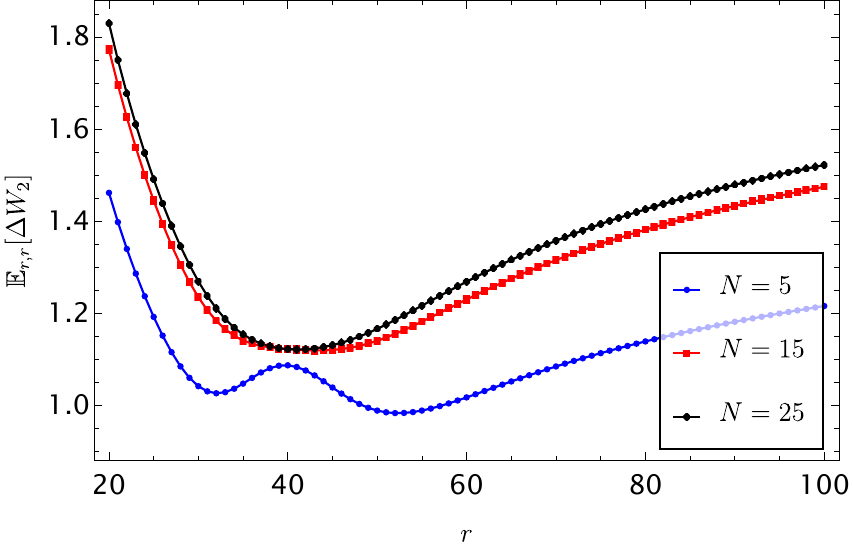}
    \caption{Expected wealth difference $\mathbb{E}_{r,r}[\Delta W_2]$ as a function of $r$, the length of the portion of string offered for sale with a fixed number of Heads, $h=20$.}
    \label{fig:Err}
\end{figure}

The calculation is performed in Appendix \ref{app:xy_buyer_exp} and gives eventually
\begin{equation}   \mathbb{E}_{r,r}\left[\frac{\delta_{\sigma_t,\sigma^{(2)}_{t}}}{N_t}\right]=C_N X_r(h)+D_N Y_r(h)\ ,
\label{eq:Err}
\end{equation}
with the coefficients $C_N$ and $D_N$ defined in Eqs.~\eqref{eq:Cn} and \eqref{eq:Dn}, and $X_{r}$ and $Y_{r}$ given by Eqs.~\eqref{eq:Xrapp} and \eqref{eq:Yrapp} respectively.
Using this result, the expected wealth difference $\mathbb{E}_{r,r}[\Delta W_2]$ appearing in \eqref{psimax} can be easily computed from Eq.~\eqref{DeltaWi}. This expectation is shown in Fig. \ref{fig:Err}, for various numbers of players, corresponding to values lower (blue curve) and higher (red and black curves), than the critical number $N^\star$ of players, respectively, and fixed number of observed Heads $h=20$. 
The expectation is higher for strongly biased coins, i.e. for
small and high values of $r$, respectively. As in the limits 
$r\to h$ and $r \gg h$ the bias is equally strong (albeit opposite), the curve looks stretched in the region $r>h/2$ compared to the region $r<h/2$.

\section{Conclusions and Outlook}\label{sec:conclusion}
We have examined the decision-making process of a set of agents in the context of a broad range of stochastic games in which some ``players'' may possess valuable information about previous realizations of the game. We propose a statistical physics framework to tackle the problem of pricing information exchange and explore the effect of information asymmetry on the players' pay-off. We thus offer a guideline to the reader interested in applying the framework to a particular game, emphasizing on the necessary modelling decisions required.

We have then specialized our analysis to a simple binary-outcome game of chance, where a fixed number $N$ of players pay a fee $\phi$ 
(which can be set, without loss of generality, to one) 
to be able to independently place a bet on one of the two possible outcomes of an unknown underlying stochastic process, which they model as a biased coin. At each round, all (if any) players placing a successful bet will split the entry fees pot as pay-off. We assume that none of the players is aware of the actual ``coin'' bias (or whether the process is indeed a coin toss), but will estimate it based on the information they hold about past outcomes of the game.

In particular, player 1 (the \emph{data holder}) is assumed to hold a long string of $R\gg 1$ past outcomes of the coin toss and number $H$ of Heads showing therein, which they can either decide to use exclusively to improve their betting performance, or partially share with player 2 in exchange for a fee.

We have determined analytically the optimal price curves for the seller and the buyer for a single round of the game, which determine the fair price range the data holder should put their information up for sale, as a function of $N,R,H$ and the length $r$ of the sub-string of data offered for sale and the number $h$ of Heads therein. 

Our results show that this game exhibits a rich 
behavior characterized by a sharp phase transition as the number of players is tuned across a critical value $N^\star$, which depends on the other informational parameters. Above this critical value, i.e. $N\geq N^\star$, the transaction is always appealing to the data holder: the reason is that in a crowded game, there are enough resources -- the entry fees paid by the $N-2$ uninformed players -- to make the rigged game always profitable at least for the seller. The prospective buyer could also benefit from the transaction in the \emph{symbiotic} regimes, but may actually leave themselves vulnerable to exploitation by a bad-faith seller in the \emph{``prey-predator''} regime, where the quality of data being offered is low enough that it would actually mislead the buyer into betting on the ``wrong'' outcome. On the other hand, below the critical number of players, i.e. $N<N^\star$, it is no longer true that the transaction is always profitable for the seller and the \emph{competition} for scarce resources may make it more convenient for the seller to hold on to the data and not share any information with the prospective buyer. 
Therefore, the ecology of the ``information landscape'' of this game is incredibly rich, even though we have restricted our study to arguably one of the simplest settings (i.e. binary outcomes, memoryless processes, and a single pairwise data transaction being arranged).

From a general point of view, our work provides a novel framework, grounded in complexity science, to address a crucial and understudied problem in econometrics and in the emerging field of \emph{infonomics}, which concerns quantifying the price of information. In particular, we hope that our study can lay the foundations for further work aimed at enabling a fair assignment of monetary value to an asset as intangible as information, in a controlled and replicable setting, in absence of a data exchange.

The framework we proposed is very general and flexible, and it could accommodate several modifications aimed at improving the realism of the setting. 

For example, we have not considered here the ``improved'' knowledge about the game that all players gain as more and more rounds are played. One could imagine that the information about the first $M$ outcomes will feed into the players' $M+1$-th bet, generating interesting \emph{memory effects}. Similarly, one could study the effect of revealing a small ``preview'', free of charge, of the data to one or more potential buyers on the seller's profit, irrespective of whether any transaction would occur.

Another possible outlook would be to assume that the length $R$ of the string of past outcomes held by one player is of finite (possibly short) length. This would allow for a further interesting effect to explore, namely the possibility that the data be corrupted or unreliable to begin with.

In addition, the number of players at each round can be made random, and different levels of fees may be imposed, corresponding to different privileges (e.g. the possibility of placing multiple bets in a multi-outcome game). Similarly, agents could randomly decide to pass (skip rounds) or place bets with a probability proportional to the strength of the bias, should they hold any extra information.

We may also consider different generating mechanisms for the outcome altogether, with or without memory (e.g. Markov chains), as well as more refined betting strategies than the simple uniform prior associated with a constant (and identical for all players) $g(\varrho)$. 

Also, the model can be generalized to games with multiple outcomes, several players holding different degrees of information (e.g. past time series of different lengths), and including multiple pairwise transactions between players -- for instance on an underlying network structure of interactions -- which are likely to create interesting correlations and feedback loops among the players.

In addition, we have not yet discussed any price implication that may stem from establishing a disclosure policy. For example, a more refined version of the game might involve the introduction of ``data scrambling'' mechanisms when the offer of the transaction is made by the data holder. This setting would allow the prospective buyer to get a glimpse of the quality of data held by the seller, without of course receiving the full information before the price is agreed upon.

We may also consider the full distribution of the wealth changes (not just their averages) when constructing the utility functions for the players or extend risk-neutral pricing to risk-averse or risk-seeking agents. Furthermore, our model is well-placed to further explore the most peculiar features of information assets. For instance, one may consider the situation where the buyer re-sells part of the data they bought to a third party, or where there is a ``data breach'' so that some players may get hold of part of the information during the exchange process. Alternatively, the seller may exploit the non-rival nature of the asset and sell (possibly different) sub-strings to multiple players. These modifications would likely lead to richer feedback effects.

In summary, the framework we proposed lends itself to a number of interesting generalizations and extensions, where the intangible nature of the asset being exchanged is assigned a monetary value on the basis of an expected utility maximization approach, with interesting interpretations in the context of the ecology of the game and resource sharing thereof.

\begin{acknowledgments}
P.V. acknowledges support from UKRI Future Leaders Fellowship Scheme
(No. MR/S03174X/1).
\end{acknowledgments}


%


\newpage
\section*{Appendix}
\appendix

\section{Optimal selection of the betting strategy}\label{app:strategy}

We consider here the problem of selecting the best betting parameter $g(\varrho)$ appearing in Eq.~\eqref{eq:strategygzero}, based on the best estimate $\varrho$ of the actual coin bias made by a player having some information $I$ about the game. 

Natural choices for $g(\varrho)$ could be $g(\varrho)=\varrho$ and $g(\varrho)=\theta(\varrho-1/2)$, with $\theta(x)$ the Heaviside step function. In the former case [$g(\varrho)=\varrho$], the player will choose to bet Heads or Tails with the same probability they have estimated for the actual bias of the coin -- so if they think that the coin has a $75\%$ bias in favor of Heads, they will bet $75\%$ of the times on Head and $25\%$ of the times on Tail. In the latter case, the player will \emph{always} bet in the direction of the bias (however strong it is), so in the example above of a coin whose bias is estimated at $75\%$ Heads, the player will always bet Heads. It can be shown that this second choice leads to a better outcome on average.

Indeed, if the player's $j$ best estimate of the pdf of the coin bias parameter is $f_I(\varrho)$, and they play according to the strategy in Eq.~\eqref{eq:strategygzero}, their probability of winning (averaged over many draws of the bias parameter) is given by
\begin{equation}
    p_j = \int_0^1 d\varrho~f_I(\varrho) [\varrho g(\varrho)+(1-\varrho)(1-g(\varrho))]\ ,
    \label{eq:winning_prob}
\end{equation}
where the first term in square brackets is the probability that the coin comes up Heads and the player has bet Heads, and the second term is the probability that the coin comes up Tails and the player has bet Tails. It is a trivial exercise to show that $p_j=2/3$ if $g(\varrho)=\varrho$ and $p_j=3/4$ if $g(\varrho)=\theta(\varrho-1/2)$ (for the case of uniform prior $f_I(\varrho)=1$). We can generalize this result by rewriting the integral in Eq.~\eqref{eq:winning_prob} as follows
\begin{align}
p_j &= \int_0^1 d\varrho~f_I(\varrho)  (1-\varrho) + \int_0^1 d\varrho~f_I(\varrho)  g(\varrho)(2\varrho-1)\nonumber\\
\nonumber    &= 1-\left<\varrho\right> + \int_{1/2}^{1} d\varrho~f_I^{(j)}(\varrho)  g(\varrho)(2\varrho-1)\\
    &- \int_0^{1/2} d\varrho~f_I(\varrho)  g(\varrho)(1-2\varrho)\ ,
\end{align}
where the first part of the expression is a constant, and the remaining integrals contain only non-negative terms. Now we can see that the choice $ g(\varrho)=\theta(\varrho-1/2)$ (irrespective of the amount of information $I$) both minimizes the negative and maximizes the positive contribution to the sum so that for \emph{any} $g_I(\varrho)$ with support and range in $[0,1]$ it holds that
\begin{equation}
    \int_0^1 d\varrho~f_I(\varrho)  g(\varrho)(2\varrho-1) \leq \int_{1/2}^{1} d\varrho~f_I(\varrho)  g(\varrho)(2\varrho-1)\ .
\end{equation}
This inequality justifies the choice in Eq.~\eqref{eq:bettingg}. A completely analogous argument holds in the case where the true bias of the coin $\rho$ is known.
\vspace{20pt}

\section{Calculation of $\mathbb{E}_{0}$ and $\mathbb{E}_{0,\cancel{\rho}}$}\label{app:E0}

\subsection{Average $\mathbb{E}_{0}$}\label{app:E0rho}

Assume now that no information is available to any player, but we look at the expected wealth difference of Player 2 from the point of view of an omniscient observer, who knows that the coin tosses are indeed independent and biased, with bias $\rho$. Using the identity $\frac{1}{x} = \int_0^\infty ds~ e^{-sx}$, for $x>0$ to lift the denominator $N_t$, we can write (exploiting sum factorization)
\begin{widetext}
\begin{align}
\nonumber \mathbb{E}_{0}\left[\frac{\delta_{\sigma_t,\sigma^{(2)}_{t}}}{N_t}\right] &=\int_0^\infty ds \int_0^1 d\rho_1~f_0(\rho_1)\cdots\int_0^1 d\rho_N~f_0(\rho_N)\sum_{\sigma_t=\pm 1} P_{\rho}(\sigma_t)\sum_{\sigma_t^{(1)}=\pm 1}\tilde P_{\rho_1}(\sigma_t^{(1)})\times\\
&\times e^{-s \delta_{\sigma_t,\sigma_t^{(1)}}} \sum_{\sigma_t^{(2)}} \tilde P_{\rho_2}(\sigma_t^{(2)}) \delta_{\sigma_t,\sigma_t^{(2)}}e^{-s \delta_{\sigma_t,\sigma_t^{(2)}}}\cdots \sum_{\sigma_t^{(N)}} \tilde P_{\rho_N}(\sigma_t^{(N)})e^{-s \delta_{\sigma_t,\sigma_t^{(N)}}}=
\label{eq:Ezerorhobase}\\
&=\int_0^\infty ds \int_0^1 d\rho_2~f_0(\rho_2) [\rho g(\rho_2) I_+(N-1,s)+ (1-\rho)(1-g(\rho_2)) I_-(N-1,s)]\ ,\label{eq:finalexpnoinforho}
\end{align}
\end{widetext}
where
\begin{align}
\nonumber   & I_+ (N-1,s) =\\
\nonumber    &=e^{-s}\prod_{k\neq 2}^N \left[\int_0^1 d\varrho~ f_0(\varrho) \sum_{\sigma=\pm 1} \tilde P_\varrho (\sigma)\exp(-s \delta_{\sigma,+1})\right]\\
    &=e^{-s}\prod_{k\neq 2}^N \left[\int_0^1 d\varrho~ f_0(\varrho) [g(\varrho) e^{-s}+(1-g(\varrho))]\right]\ , \label{eq:Ip_N-2}
\end{align}
and
\begin{align}
\nonumber & I_-(N-1,s) =\\
\nonumber     &= e^{-s}\prod_{k\neq 2}^N \left[\int_0^1 d\varrho~ f_0(\varrho) \sum_{\sigma=\pm 1} \tilde P_\varrho (\sigma)\exp(-s \delta_{\sigma,-1})\right]\\
    &= e^{-s}\prod_{k\neq 2}^N \left[\int_0^1 d\varrho~ f_0(\varrho) [g(\varrho) +(1-g(\varrho))e^{-s}]\right]\ .
    \label{eq:Im_N-2}
\end{align}
Using 
$f_0(\rho_2)=1$, $g(\rho_2)=\theta(\rho_2-1/2)$ and the symmetry condition $1-g(\rho_2)=g(1-\rho_2)$ (which further implies $\int_0^1 d\rho_2~g(\rho_2)=\int_0^1 d\rho_2 (1-g(\rho_2)$), we get that the $\rho$ dependence drops out, and 
\begin{align}
\nonumber\mathbb{E}_{0} &\left[\frac{\delta_{\sigma_t,\sigma^{(2)}_{t}}}{N_t}\right]   =
 \int_0^\infty ds~e^{-s}\left(\frac{1}{2}e^{-s} + \frac{1}{2}\right)^{N-1} \times\\
 &\times\int_0^1 d\rho_2 g(\rho_2)= \frac{2-2^{1-N}}{N}\int_0^1 d\rho_2 g(\rho_2)\ ,
\end{align}
independently of $\rho$. Computing the integral explicitly, leads to \eqref{eq:true_baseline},
showing that the expectation is always negative. This is expected as -- in the event of all players betting on the wrong outcome -- the ``dealer'' would keep the total entry fee collected from the players, which would not be redistributed among them.

In the next subsection, we perform the same calculation but this time assuming the viewpoint of player 2, who does not know the actual value of the coin bias $\rho$ and will therefore use their own estimate $\rho_2$ in lieu of $\rho$.

\subsection{Average $\mathbb{E}_{0,\cancel{\rho}}$}

Here we compute the expected wealth change, from the point of view of player 2 (or, equivalently, any other uninformed player),  using the estimate $\rho_2$ to replace the true, unknown, coin bias $\rho$. We refer to this expectation as  $\mathbb{E}_{0,\cancel{\rho}}[\Delta W_2]$ to remark the use of a proxy for $\rho$ in the computation of \eqref{eq:expzerorho1}, rather than leaving it as a free parameter, even though it eventually drops out. The computation is carried out as before, starting from
\begin{widetext}
\begin{align}
\nonumber \mathbb{E}_{0,\cancel{\rho}}\left[\frac{\delta_{\sigma_t,\sigma^{(2)}_{t}}}{N_t}\right] &=\int_0^\infty ds \int_0^1 d\rho_1~f_0(\rho_1)\cdots\int_0^1 d\rho_N~f_0(\rho_N)\sum_{\sigma_t=\pm 1} P_{\rho_2}(\sigma_t)\sum_{\sigma_t^{(1)}=\pm 1}\tilde P_{\rho_1}(\sigma_t^{(1)})\times\\
&\times e^{-s \delta_{\sigma_t,\sigma_t^{(1)}}} \sum_{\sigma_t^{(2)}} \tilde P_{\rho_2}(\sigma_t^{(2)}) \delta_{\sigma_t,\sigma_t^{(2)}}e^{-s \delta_{\sigma_t,\sigma_t^{(2)}}}\cdots \sum_{\sigma_t^{(N)}} \tilde P_{\rho_N}(\sigma_t^{(N)})e^{-s \delta_{\sigma_t,\sigma_t^{(N)}}}=\label{eq:int1}
\\
&=\int_0^\infty ds \int_0^1 d\rho_2~f_0(\rho_2) [\rho_2 g(\rho_2) I_+(N-1,s)+ (1-\rho_2)(1-g(\rho_2)) I_-(N-1,s)]\ ,\label{eq:int2}
\end{align}
\end{widetext}
where $I_\pm (N-1,s)$ are defined in \eqref{eq:Ip_N-2} and \eqref{eq:Im_N-2}.

Assuming again that all uninformed players will estimate the bias parameter of the coin uniformly in $[0,1]$, we get again for $g(\varrho)=\theta(\varrho-1/2)$ that
\begin{equation}
    I_+(N-1,s)=I_-(N-1,s)=e^{-s}\left[\frac{1}{2}e^{-s}+\frac{1}{2}\right]^{N-1}\ ,
\end{equation}
from which it follows that
\begin{align}
\nonumber\mathbb{E}_{0,\cancel{\rho}}\left[\frac{\delta_{\sigma_t,\sigma^{(2)}_{t}}}{N_t}\right] 
    &= \int_0^\infty ds~e^{-s}\left(\frac{1}{2}e^{-s} + \frac{1}{2}\right)^{N-1} \times\\
&\times\mathbb{E}_{0,\cancel{\rho}}\left[\delta_{\sigma_t,\sigma^{(2)}_{t}}\right] =    \frac{3}{4}\frac{2-2^{1-N}}{N}\ ,
    \label{eq:upside_lucky}
\end{align}
where
\begin{align}
\nonumber\mathbb{E}_{0,\cancel{\rho}}\left[\delta_{\sigma_t,\sigma^{(2)}_{t}}\right] &=   \int_0^1 d\rho_2 \left(\rho_2 g(\rho_2) + (1-\rho_2)(1-g(\rho_2))\right)\\
&=\frac{3}{4} 
\end{align}
is the winning probability estimated by player $2$, who does not have any knowledge of the actual bias of the coin. 

Taking the expectation of Eq.~\eqref{DeltaWi} and inserting Eq.~\eqref{eq:upside_lucky}, we get 
\begin{align}
\nonumber    \mathbb{E}_{0,\cancel{\rho}}\left[\Delta W_2 \right] &= -M+M \frac{3(1-2^{-N})}{2}\\
    &=\frac{1}{2}M \left(1 - \frac{3}{2^{N}}\right)\ .\label{expDeltaWzero}
\end{align}

Comparing \eqref{expDeltaWzero} and \eqref{eq:true_baseline}, we find that $\mathbb{E}_{0,\cancel{\rho}}\left[\Delta W_2 \right] >\mathbb{E}_{0}\left[\Delta W_2\right] $. This is intuitive, as in $\mathbb{E}_{0,\cancel{\rho}}$ the player is using the same parameter $\rho_2$ to estimate their best strategy \emph{and} the actual bias of the coin, which obviously leads to a more optimistic outlook on their game.

\section{Calculation of winning probability for the data holder and asymptotics for $R\to\infty$}\label{app:winning_probability}

We compute here the probability that the data holder wins the bet in a single round (estimated by the data holder themselves)  assuming that they hold a string of the past $R$ outcomes and will use it to 
place their bet

\begin{widetext}
\begin{align}
\nonumber\mathbb{E}_R\left[\delta_{\sigma_t,\sigma^{(1)}_{t}}\right]&= \int_0^1 d\rho_1~f_R(\rho_1)\int_0^1 d\rho_2~f_0(\rho_2)\cdots\int_0^1 d\rho_N~f_0(\rho_N)\sum_{\sigma_t=\pm 1} P_{\rho_1}(\sigma_t)\sum_{\sigma_t^{(1)}=\pm 1}\tilde P_{\rho_1}(\sigma_t^{(1)})\times\\
\nonumber &\times \delta_{\sigma_t,\sigma_t^{(1)}}\sum_{\sigma_t^{(2)}} \tilde P_{\rho_2}(\sigma_t^{(2)})\cdots \sum_{\sigma_t^{(N)}} \tilde P_{\rho_N}(\sigma_t^{(N)})=
\\
&= \int_0^1 d\rho_1~f_R(\rho_1) [\rho_1 g(\rho_1)+(1-\rho_1)(1-g(\rho_1)))]\ .\label{eq:finalexpnoinforhoimprovedappB}
\end{align}
\end{widetext}
Here, we have used the fact that the data holder (player 1) will estimate the actual coin bias \emph{and} the parameter appearing in their own best strategy as $\rho_1$, which is drawn from the posterior pdf $f_R(\rho_1)$ given in Eq.~\eqref{eq:seller_posterior}. We also used that the betting distributions $\tilde P_\varrho$ are normalized to unity.

Performing the elementary integral in \eqref{eq:finalexpnoinforhoimprovedappB} with $g(\rho_1)=\theta(\rho_1-1/2)$ and $f_R(\rho_1)$ given in Eq.~\eqref{eq:seller_posterior} we get

\begin{align}
\nonumber &\mathbb{E}_R\left[\delta_{\sigma_t,\sigma^{(1)}_{t}}\right]=\frac{(R+1)!}{H! (R-H)! 2^{R+2}}\times \\
\nonumber &\times \left[F(H+1,R-H)+F(R-H+1,H)\right]\\
&=:\Xi_R(H)\ ,\label{Xiappendix}
\end{align}
where $F(x,y)$ is defined in Eq.~\eqref{eq:F_definition} in terms of the $_2F_1$ hypergeometric function.

We can now compute the asymptotics $\Xi_R(\alpha R)$ for large $R$ for a fraction $0\leq\alpha\leq 1$ of Heads seen in the long string of data held by player 1. From \eqref{Xiappendix}, we need the following saddle-point asymptotics 
\begin{align}
\nonumber    F(\alpha R+1,R-\alpha R) &=\int_0^1 dt (1+t)^{\alpha R+1}(1-t)^{R(1-\alpha)}=\\
    &=\int_0^1 dt (1+t)\exp[R g_\alpha(t)]\\
\nonumber    F(R-\alpha R+1,\alpha R) &=\int_0^1 dt (1+t)^{R-\alpha R+1}(1-t)^{\alpha R}=\\
    &=\int_0^1 dt (1+t)\exp[R g_{1-\alpha}(t)]\ ,
\end{align}
with
\begin{align}
    g_\alpha(t)=\alpha \log(1+t)+(1-\alpha)\log(1-t)\ ,
\end{align}
from which 
\begin{align}
    g_\alpha'(t^\star)=0\Rightarrow t^\star = 2\alpha-1\ ,
\end{align}
which is within the integration interval for $1/2<\alpha<1$. Using the Stirling approximation for the prefactor 
\begin{align}
\nonumber   \frac{(R+1)!}{H! (R-H)! 2^{R+m}} &\sim e^{-R\left[\alpha  \log (1-\alpha )+(1-\alpha)\log (1-\alpha )+\log (2)\right]}\times\\
   &\times \frac{1}{2^m}\sqrt{\frac{R}{2\pi\alpha(1-\alpha)}}\label{stirling}
\end{align}
and computing $g_\alpha(t^\star)$, we see that the leading exponential terms cancel out exactly, and for $m=2$ combining all the prefactors together we finally obtain 
Eq.~\eqref{XiRasympt} of the main text.

\section{Calculation of $\mathbb{E}_{R,r}$}\label{app:xy_seller_exp}

We start from Eq.~\eqref{eq:ERrtext}
\begin{widetext}
\begin{align}
&\nonumber\mathbb{E}_{R,r}\left[\frac{\delta_{\sigma_t,\sigma^{(1)}_{t}}}{N_t}\right] =\int_0^\infty ds \int_0^1 d\rho_1~f_{R,r}(\rho_1)\int_0^1 d\rho_2~f_{r,r}(\rho_2)\cdots\int_0^1 d\rho_N~f_0(\rho_N)\sum_{\sigma_t=\pm 1} P_{\rho_1}(\sigma_t)\sum_{\sigma_t^{(1)}=\pm 1}\tilde P_{\rho_1}(\sigma_t^{(1)})\times\\
&\times \delta_{\sigma_t,\sigma_t^{(1)}}e^{-s \delta_{\sigma_t,\sigma_t^{(1)}}} \sum_{\sigma_t^{(2)}} \tilde P_{\rho_2}(\sigma_t^{(2)})e^{-s \delta_{\sigma_t,\sigma_t^{(2)}}}\cdots \sum_{\sigma_t^{(N)}} \tilde P_{\rho_N}(\sigma_t^{(N)})e^{-s \delta_{\sigma_t,\sigma_t^{(N)}}}=
\\
&=\int_0^\infty ds \int_0^1 d\rho_1~f_{R,r}(\rho_1)\int_0^1 d\rho_2~f_{r,r}(\rho_2) [\rho_1 g(\rho_1) I_+(N-2,s)\chi_+(\rho_2,s)+ (1-\rho_1)(1-g(\rho_1)) I_-(N-2,s)\chi_-(\rho_2,s)]\ ,\label{eq:ErrWealthSeller}
\end{align}
\end{widetext}
where $I_\pm$ are defined in Eq.~\eqref{eq:Ip_N-2} and Eq.~\eqref{eq:Im_N-2} and
\begin{align}
    \chi_+(\varrho,s) &= g(\varrho) e^{-s}+1-g(\varrho)\label{eq:chip}\\
    \chi_-(\varrho,s) &= g(\varrho)+(1-g(\varrho))e^{-s}\ .
    \label{eq:chim}
\end{align}
Here, we have used factorization of the summations after $N_t$ is lifted up using the $s$-identity, and the fact that $N-2$ summations (corresponding to the uninformed players) are identical.

Also, for uniform priors $f_0$, we have that
\begin{equation}
    I_+(N-2,s)=I_-(N-2,s)=e^{-s}\left[\frac{1}{2}e^{-s}+\frac{1}{2}\right]^{N-2}\ ,
    \label{eq:I_N-2_value}
\end{equation}
leading to
\begin{equation}   \mathbb{E}_{R,r}\left[\frac{\delta_{\sigma_t,\sigma^{(1)}_{t}}}{N_t}\right]=C_N X_{R,r}(x,h)+D_N Y_{R,r}(x,h)\ ,
\end{equation}
where
\begin{align}
    C_N &= \int_0^\infty ds~e^{-2s}\left[\frac{1}{2}e^{-s}+\frac{1}{2}\right]^{N-2}=\frac{2 N+2^{2-N}-4}{(N-1) N} \label{eq:Cn}\\
     D_N &= \int_0^\infty ds~e^{-s}\left[\frac{1}{2}e^{-s}+\frac{1}{2}\right]^{N-2}=\frac{2-2^{2-N}}{N-1} \label{eq:Dn} \ ,
\end{align} 
and
\begin{align}
 \nonumber     & X_{R,r}(x,h) =\int_0^1 d\rho_1 f_{R,r}(\rho_1)\int_0^1 d\rho_2 f_{r,r}(\rho_2)\times\\
 &\times\left[\rho_1 g(\rho_1)g(\rho_2)+(1-\rho_1)(1-g(\rho_1))(1-g(\rho_2))\right]\ ,\\
 \nonumber &    Y_{R,r}(x,h) =\int_0^1 d\rho_1 f_{R,r}(\rho_1)\int_0^1 d\rho_2 f_{r,r}(\rho_2)\times \\
 &\times\left[\rho_1 g(\rho_1)(1-g(\rho_2))+(1-\rho_1)(1-g(\rho_1))g(\rho_2)\right]\ .
\end{align}

We start from
\begin{align}
\nonumber    & X_{R,r}(x,h) =\int_0^1 d\rho_1 f_{R,r}(\rho_1)\int_0^1 d\rho_2 f_{r,r}(\rho_2)\times\\
\nonumber    &\times \left[\rho_1 g(\rho_2)g(\rho_1)+(1-\rho_1)(1-g(\rho_2))(1-g(\rho_1))\right]\\
    &=J_1+J_2\ ,
\end{align}
where
\begin{align}
    J_1 &= \int_{1/2}^1 d\rho_2 f_{r,r}(\rho_2)\int_{1/2}^1 d\rho_1 f_{R,r}(\rho_1)\rho_1\\ 
    J_2 &=\int_{0}^{1/2} d\rho_2 f_{r,r}(\rho_2)\int_{0}^{1/2} d\rho_1 f_{R,r}(\rho_1)(1-\rho_1)\ .
\end{align}

First, we have
\begin{align}
&\int_{1/2}^1 d\rho_2 f_{r,r}(\rho_2) = C_{r,h}\frac{1}{2^{r+1}}F(h,r-h)\\
&\int_{1/2}^1 d\rho_1 f_{R,r}(\rho_1)\rho_1 = \frac{C_{R,x+h}}{2^{R+2}}F(h+x+1,R-(h+x))\\
&\int_{0}^{1/2} d\rho_2 f_{r,r}(\rho_2) =C_{r,h}\frac{1}{2^{r+1}}F(r-h,h)\\
\nonumber &\int_{0}^{1/2} d\rho_1f_{R,r}(\rho_1)(1-\rho_1) =\frac{C_{R,h+x}}{2^{R+2}}\times \\
&\times F(R-(h+x)+1,h+x)\ ,
\end{align}
where $C_{r,h}=(r+1)!/(h!(r-h)!)$ and $C_{R,x+h}=(R+1)!/((h+x)!(R-(h+x))!)$. $F(x,y)$ is defined in Eq.~\eqref{eq:F_definition}.

Therefore
\begin{align}
\nonumber   X_{R,r}(x,h) &=\frac{C_{R,h+x}C_{r,h} }{2^{R+r+3}}\times\\
\nonumber   &\times \left\{F(h,r-h)F(h+x+1,R-(h+x))\right.\\
   &\left.+F(r-h,h)F(R-(h+x)+1,h+x)\right\}\ .
   \label{eq:XRrapp}
\end{align}
Similarly,
\begin{align}
 \nonumber  & Y_{R,r}(x,h) =\int_0^1 d\rho_1 f_{R,r}(\rho_1)\int_0^1 d\rho_2 f_{r,r}(\rho_2)\times\\
\nonumber &\times\left[\rho_1 g(\rho_1)(1-g(\rho_2))+(1-\rho_1)(1-g(\rho_1))g(\rho_2)\right]\\
 &=K_1+K_2\ ,
\end{align}
where
\begin{align}
    K_1 &= \int_{0}^{1/2} d\rho_2 f_{r,r}(\rho_2)\int_{1/2}^1 d\rho_1 f_{R,r}(\rho_1)\rho_1\\ 
    K_2 &=\int_{1/2}^{1} d\rho_2 f_{r,r}(\rho_2)\int_{0}^{1/2} d\rho_1 f_{R,r}(\rho_1)(1-\rho_1)\ .
\end{align}
Using the previously computed elementary integrals, we can immediately write
\begin{align}
\nonumber   Y_{R,r}(x,h) &=\frac{C_{R,h+x} C_{r,h}}{2^{R+r+3}}\times\\
\nonumber   &\times\left\{F(r-h,h)F(h+x+1,R-(h+x))\right.\\
   &\left.+F(h,r-h)F(R-(h+x)+1,h+x)\right\}\ . 
   \label{eq:YRrapp}
\end{align}

\section{Calculation of $\mathbb{E}_{r,r}$}\label{app:xy_buyer_exp}

We start from Eq.~\eqref{eq:ErrWealthBuyer}
\begin{widetext}
\begin{align}
&\nonumber\mathbb{E}_{r,r}\left[\frac{\delta_{\sigma_t,\sigma^{(2)}_{t}}}{N_t}\right] =\int_0^\infty ds \int_0^1 d\rho_1~f_{r,r}(\rho_1)\int_0^1 d\rho_2~f_{r,r}(\rho_2)\cdots\int_0^1 d\rho_N~f_0(\rho_N)\sum_{\sigma_t=\pm 1} P_{\rho_2}(\sigma_t)\sum_{\sigma_t^{(2)}=\pm 1}\tilde P_{\rho_2}(\sigma_t^{(2)})\times\\
&\times \delta_{\sigma_t,\sigma_t^{(2)}}e^{-s \delta_{\sigma_t,\sigma_t^{(2)}}} \sum_{\sigma_t^{(1)}} \tilde P_{\rho_1}(\sigma_t^{(1)})e^{-s \delta_{\sigma_t,\sigma_t^{(1)}}}\cdots \sum_{\sigma_t^{(N)}} \tilde P_{\rho_N}(\sigma_t^{(N)})e^{-s \delta_{\sigma_t,\sigma_t^{(N)}}}=
\\
&=\int_0^\infty ds \int_0^1 d\rho_1~f_{r,r}(\rho_1)\int_0^1 d\rho_2~f_{r,r}(\rho_2) [\rho_2 g(\rho_2) I_+(N-2,s)\chi_+(\rho_1,s)+ (1-\rho_2)(1-g(\rho_2)) I_-(N-2,s)\chi_-(\rho_1,s)]\ ,\label{eq:ErrWealthBuyer}
\end{align}
\end{widetext}
where $\chi_\pm$ is defined in Eq.~\eqref{eq:chip} and Eq.~\eqref{eq:chim}, and $I_\pm$ are defined in Eq.~\eqref{eq:Ip_N-2} and Eq.~\eqref{eq:Im_N-2}. Also, using Eq.~\eqref{eq:I_N-2_value}, we have that
\begin{equation}   \mathbb{E}_{r,r}\left[\frac{\delta_{\sigma_t,\sigma^{(2)}_{t}}}{N_t}\right]=C_N X_r(h)+D_N Y_r(h)\ ,
\end{equation}
where $C_N$ and $D_N$ are defined respectively in Eq.~\eqref{eq:Cn} and Eq.~\eqref{eq:Dn}, also

\begin{align}
    \nonumber & X_r(h) =\int_0^1 d\rho_1 f_{r,r}(\rho_1)\int_0^1 d\rho_2 f_{r,r}(\rho_2)\times \\
    &\times\left[\rho_2 g(\rho_2)g(\rho_1)+(1-\rho_2)(1-g(\rho_2))(1-g(\rho_1))\right]\ ,\\
\nonumber &     Y_r(h) =\int_0^1 d\rho_1 f_{r,r}(\rho_1)\int_0^1 d\rho_2 f_{r,r}(\rho_2)\times\\
&\times \left[\rho_2 g(\rho_2)(1-g(\rho_1))+(1-\rho_2)(1-g(\rho_2))g(\rho_1)\right]\ .
\end{align}

We start from
\begin{align}
   \nonumber & X_r(h) =\int_0^1 d\rho_1 f_{r,r}(\rho_1)\int_0^1 d\rho_2 f_{r,r}(\rho_2)\times\\
   \nonumber &\times\left[\rho_2 g(\rho_2)g(\rho_1)+(1-\rho_2)(1-g(\rho_2))(1-g(\rho_1))\right]\\
   &=J_1'+J_2'\ ,
\end{align}
where
\begin{align}
    J_1' &= \int_{1/2}^1 d\rho_1 f_{r,r}(\rho_1)\int_{1/2}^1 d\rho_2 f_{r,r}(\rho_2)\rho_2\\ 
    J_2' &=\int_{0}^{1/2} d\rho_1 f_{r,r}(\rho_1)\int_{0}^{1/2} d\rho_2 f_{r,r}(\rho_2)(1-\rho_2)\ .
\end{align}
First, we have
\begin{align}
&\int_{1/2}^1 d\rho_1 f_{r,r}(\rho_1) = C_{r,h}\frac{1}{2^{r+1}}F(h,r-h)\\
&\int_{1/2}^1 d\rho_2 f_{r,r}(\rho_2)\rho_2 = C_{r,h}\frac{1}{2^{r+2}}F(h+1,r-h)\\
&\int_{0}^{1/2} d\rho_1 f_{r,r}(\rho_1) =C_{r,h}\frac{1}{2^{r+1}}F(r-h,h)\\
&\int_{0}^{1/2} d\rho_2 f_{r,r}(\rho_2)(1-\rho_2) =C_{r,h}\frac{1}{2^{r+2}}F(r-h+1,h)\ ,
\end{align}
where $C_{r,h}=(r+1)!/(h!(r-h)!)$ and $F(x,y)$ is defined in Eq.~\eqref{eq:F_definition}.

Therefore,
\begin{align}
\nonumber   X_r(h) &=(C_{r,h})^2 \frac{1}{2^{2r+3}}\left\{F(h,r-h)F(h+1,r-h)\right.\\
   &\left. +F(r-h,h)F(r-h+1,h)\right\}\ . 
   \label{eq:Xrapp}
\end{align}
Similarly,
\begin{align}
    \nonumber & Y_r(h) =\int_0^1 d\rho_1 f_{r,r}(\rho_1)\int_0^1 d\rho_2 f_{r,r}(\rho_2)\times\\
    \nonumber &\times\left[\rho_2 g(\rho_2)(1-g(\rho_1))+(1-\rho_2)(1-g(\rho_2))g(\rho_1)\right]\\
    &=K_1'+K_2'
\end{align}
where
\begin{align}
    K_1' &= \int_{0}^{1/2} d\rho_1 f_{r,r}(\rho_1)\int_{1/2}^1 d\rho_2 f_{r,r}(\rho_2)\rho_2\\ 
    K_2' &=\int_{1/2}^{1} d\rho_1 f_{r,r}(\rho_1)\int_{0}^{1/2} d\rho_2 f_{r,r}(\rho_2)(1-\rho_2)\ .
\end{align}
Using the previously computed elementary integrals, we can immediately write
\begin{align}
 \nonumber  Y_r(h) &=(C_{r,h})^2 \frac{1}{2^{2r+3}}\left\{F(r-h,h)F(h+1,r-h)\right.\\
   &\left.+F(h,r-h)F(r-h+1,h)\right\}\ . 
   \label{eq:Yrapp}
\end{align}

\section{Explicit expressions for $\Psi_{\mathrm{min}}$ and $\Psi_{\mathrm{max}}$ for $M=1$}\label{app:LimitingPrice}

Let us put ourselves in the simplified setting $M=1$ (single round of the game). Recalling \eqref{DeltaWi} and the various intermediate results
\begin{align}
\mathbb{E}_{R}\left[\frac{\delta_{\sigma_t,\sigma^{(1)}_{t}}}{N_t}\right]
    &=\frac{2-2^{1-N}}{N} \Xi_R(H)\\
\mathbb{E}_{R,r}\left[\frac{\delta_{\sigma_t,\sigma^{(1)}_{t}}}{N_t}\right] &=C_N X_{R,r}(x,h)+D_N Y_{R,r}(x,h)\\
\mathbb{E}_{r,r}\left[\frac{\delta_{\sigma_t,\sigma^{(2)}_{t}}}{N_t}\right] &=C_N X_r(h)+D_N Y_r(h)\\
\mathbb{E}_0\left[\frac{\delta_{\sigma_t,\sigma^{(2)}_{t}}}{N_t}\right] 
    &=    \frac{1}{2}\frac{2-2^{1-N}}{N}\ ,
\end{align}
we get
\begin{align}
\nonumber\Psi_{\mathrm{min}}/\phi &=\max[(2-2^{1-N})\Xi_R(h+x)\\
&-N C_N X_{R,r}(x,h)-N D_N Y_{R,r}(x,h),0]\\
\nonumber\Psi_{\mathrm{max}}/\phi &=\max[N C_N X_{r}(h)+N D_N Y_{r}(h)\\
&-\frac{1}{2}(2-2^{1-N}),0]\ ,
\end{align}
where
\begin{align}
\nonumber  \Xi_R(H) &=\frac{(R+1)!}{H! (R-H)! 2^{R+2}}\left[F(H+1,R-H)+\right.\\
  &\left. +F(R-H+1,H)\right]\ ,
\end{align}
with
\begin{equation} 
F(x,y)=\int_0^1 dt~(1+t)^x (1-t)^y=\frac{\, _2F_1(1,-x;y+2;-1)}{y+1}\ .
  \label{eq:F_definitionapp}
 \end{equation}
Furthermore, we have
\begin{align}
 C_N &=\frac{2 N+2^{2-N}-4}{(N-1) N} \label{eq:Cnapp}\\
     D_N &=\frac{2-2^{2-N}}{N-1} \label{eq:Dnapp}\\
     \nonumber   X_r(h) &=(C_{r,h})^2 \frac{1}{2^{2r+3}}\left\{F(h,r-h)F(h+1,r-h)\right.\\
   &\left. +F(r-h,h)F(r-h+1,h)\right\}\\
 \nonumber  Y_r(h) &=(C_{r,h})^2 \frac{1}{2^{2r+3}}\left\{F(r-h,h)F(h+1,r-h)\right.\\
   &\left.+F(h,r-h)F(r-h+1,h)\right\}\\
   \nonumber   X_{R,r}(x,h) &=\frac{C_{R,h+x}C_{r,h} }{2^{R+r+3}}\times\\
\nonumber   &\times \left\{F(h,r-h)F(h+x+1,R-(h+x))\right.\\
   &\left.+F(r-h,h)F(R-(h+x)+1,h+x)\right\}\\
   \nonumber   Y_{R,r}(x,h) &=\frac{C_{R,h+x} C_{r,h}}{2^{R+r+3}}\times\\
\nonumber   &\times\left\{F(r-h,h)F(h+x+1,R-(h+x))\right.\\
   &\left.+F(h,r-h)F(R-(h+x)+1,h+x)\right\} 
\end{align}

in terms of constants $C_{p,q}=(p+1)!/(q!(p-q)!)$.

\end{document}